\newcommand{\pa}{\partial}
\newcommand{\be}{\begin{equation}}
\newcommand{\ee}{\end{equation}}
\newcommand{\bea}{\begin{eqnarray}}
\newcommand{\eea}{\end{eqnarray}}

\newcommand{\bt}[1]{{\bar t}}
\documentclass[11pt]{article}
\usepackage{amsmath,amsfonts,epsfig,color,latexsym}
\usepackage{amssymb}
\def\xxx#1           {{\sf hep-th/#1} }

\def\npb#1(#2)#3     {Nucl. Phys. {\bf B#1} (#2) #3 }
\def\rep#1(#2)#3     {Phys. Rept.{\bf #1} (#2) #3 }
\def\pla#1(#2)#3     {Phys. Lett. {\bf #1A} (#2) #3 }
\def\plb#1(#2)#3     {Phys. Lett. {\bf #1B} (#2) #3 }
\def\prl#1(#2)#3     {Phys. Rev. Lett.{\bf #1} (#2) #3 }
\def\prd#1(#2)#3     {Phys. Rev. {\bf D#1} (#2) #3 }
\def\ap#1(#2)#3      {Ann. Phys. {\bf #1} (#2) #3 }
\def\rmp#1(#2)#3     {Rev. Mod. Phys. {\bf #1} (#2) #3 }
\def\cmp#1(#2)#3     {Comm. Math. Phys. {\bf #1} (#2) #3 }
\def\mpla#1(#2)#3    {Mod. Phys. Lett. {\bf A#1} (#2) #3 }
\def\ijmp#1(#2)#3    {Int. J. Mod. Phys. {\bf A#1} (#2) #3 }
\def\cqg#1(#2)#3     {Class. Quant. Grav. {\bf #1} (#2) #3 }
\def\am#1(#2)#3      {Adv. Math. {\bf #1} (#2) #3 }
\def\im#1(#2)#3      {Invent. Math. {\bf #1} (#2) #3 }
\def\jhep#1(#2)#3    {JHEP {\bf #1}(#2) #3 }
\def\npps#1(#2)#3    {Nucl.Phys.Proc.Suppl. {\bf #1}(#2) #3 }
\def\jgp#1(#2)#3     {J. Geom. Phys. {\bf #1}(#2) #3 }
\def\atmp#1(#2)#3    {Adv. Theor. Math. Phys. {\bf #1} (#2) #3}
\def\lmp#1(#2)#3     {Lett. Math. Phys. {\bf #1} (#2) #3}

\begin{document}
\thispagestyle{empty}
\null\vskip-24pt \hfill AEI-2002-040 \\
\vskip-10pt \hfill  CERN-TH/2002-115  \\
\vskip-10pt \hfill  LAPTH-915/02 \\
 \vskip-10pt \hfill {\tt hep-th/0205270}
\vskip0.2truecm
\begin{center}
\vskip 0.2truecm {\Large\bf
\Large{ Conformal fields in the pp-wave limit
 }
}\\
\vskip 1truecm
{\bf G. Arutyunov$^{*,**}$\footnote{email:{\tt agleb@aei-potsdam.mpg.de}},
E. Sokatchev$^{\ddagger,\ddagger\ddagger}$
\footnote{email:{\tt Emery.Sokatchev@cern.ch} \\
$^{**}$On leave of absence from Steklov Mathematical Institute, Gubkin str.8,
117966, Moscow, Russia\\
$^{\ddagger\ddagger}$On leave of absence from Laboratoire d'Annecy-le-Vieux de
Physique Th\'{e}orique  LAPTH, B.P. 110, F-74941 Annecy-le-Vieux et l'Universit\'{e} de
Savoie      }
}\\

\vskip 0.4truecm
$^{*}$ {\it Max-Planck-Institut f\"ur Gravitationsphysik,
Albert-Einstein-Institut, \\
Am M\"uhlenberg 1, D-14476 Golm, Germany}\\
\vskip .2truecm $^{\ddagger}$ {\it CERN Theoretical Division, CH12-11 Geneva
23, Switzerland
} \\
\end{center}

\vskip 1truecm \Large
\centerline{\bf Abstract} \normalsize The pp-wave (Penrose limit) in conformal
field theory can be viewed as a special contraction of the unitary
representations of the conformal group. We study the kinematics of conformal
fields in this limit in a geometric approach where the effect of the
contraction can be visualized as an expansion of space-time. We discuss the two
common models of space-time as carrier spaces for conformal fields: One is the
usual Minkowski space and the other is the coset of the conformal group over
its maximal compact subgroup. We show that only the latter manifold and the
corresponding conformal representation theory admit a non-singular contraction
limit. We also address the issue of correlation functions of conformal fields
in the pp-wave limit. We show that they have a well-defined contraction limit
if their space-time dependence merges with the dependence on the coordinates of
the R symmetry group. This is a manifestation of the fact that in the limit the
space-time and R symmetries become indistinguishable. Our results might find
applications in actual calculations of correlation functions of composite
operators in ${\cal N}=4$ super Yang-Mills theory.

\newpage
\setcounter{page}{1}\setcounter{footnote}{0}

\section{Introduction}

${\cal N}=4$ super Yang-Mills theory is a non-trivial interacting
superconformal field theory which is believed to possess a holographic (dual)
description in terms of type IIB string theory on an $AdS_5\times S^5$
background \cite{AGMOO}. Apart from $AdS_5\times S^5$ and flat Minkowski space,
there exists another maximally supersymmetric background referred to as the
``pp-wave solution" \cite{BFHP1}. This background also arises as the Penrose
limit of the $AdS_5\times S^5$ space \cite{BFHP2}. Thus, it is of obvious
interest to understand what happens to the AdS/CFT duality in the Penrose
limit.

One of the nice features of the pp-wave solution is that the spectrum of a free
string propagating in this background can be found exactly \cite{M,MT}. The
knowledge of the spectrum offers an opportunity to study the duality conjecture
beyond the supergravity approximation. Recently, an interesting proposal
\cite{BMN} has been made to identify certain classes of gauge invariant
composite operators in ${\cal N}=4$ theory with states of the supergravity and
string theory in the pp-wave background.

Both the kinematical and dynamical information about any gauge theory is
encoded in the correlation functions of gauge invariant composite operators.
Such operators form unitary irreducible representations (UIRs) of the
superconformal algebra PSU(2,2$|$4) and the lowest bosonic components of the
corresponding supermultiplets are specified by their conformal dimension
$\Delta$, the two Lorentz spins of SO(3,1) and by a finite-dimensional UIR of
the R symmetry group SU(4). On the gauge theory side the Penrose limit consists
in taking the rank $N$ of the gauge group SU($N$) to infinity, while keeping the
Yang-Mills coupling constant $g$ fixed. According to the proposal of Ref.
\cite{BMN}, the operators surviving in this limit have a very large conformal
dimension $\Delta\sim \sqrt{N}$ and a very large U(1) charge $J$ (where U(1) is
a subgroup of the R symmetry group), such that the ratio $J^2/N$ and the
difference $\Delta-J$ are kept finite.
Some interesting perturbative
computations of two- and three-point correlation functions in ${\cal N}=4$
theory have been done recently \cite{KPSS,GMR,CFHMMPS} and they provide further
evidence that the well-defined perturbative parameter of the gauge theory in
this limit is $g^2N/J^2$.

However, we feel that the present understanding of the basic principles of the
conformal field theory (CFT) in this limit is not yet complete. Above all, this
concerns the precise definition of the Hilbert space of states and  of the
correlation functions. Indeed, the operators we are interested in have infinite
canonical dimension and therefore the space-time dependence of their
correlation functions requires a special treatment. In particular, one would
like to understand the space-time manifold where the dual gauge theory lives.
Some of these questions have already been addressed in the current literature
\cite{Gomez,Nastase,Leigh}.

The pp-wave limit can be viewed as the passage from the superconformal algebra
PSU(2,2$|$4) to the symmetry superalgebra of the pp-wave background. The
bosonic part of the latter is the semi-direct sum of h(8) with the external
automorphism algebra so(4)$\; \oplus\; $so(4)$\;\\ \oplus\; $so(2). Here h(8)
denotes two copies of the four-dimensional Heisenberg algebra h(4) with a
common central element. This passage is just a generalized In\"on\"u-Wigner
contraction \cite{BFHP1, HKS}. The contraction parameter is related to the
infinite conformal dimension of the fields.

The conformal fields are UIRs of the conformal group. Therefore, one can try to
carry out the contraction procedure directly in the Hilbert space of conformal
fields in order to see which states survive in the contraction limit and give
rise to the Hilbert space of states of the contracted group. Studying conformal
kinematics in the pp-wave limit is the aim of the present paper.

The main problem that one is confronted with is the infinite conformal
dimension $\Delta$ of the fields. We propose a mechanism in which this infinite
weight is compensated by another infinite quantum number carried by the fields,
a charge $J$ coming from the R symmetry sector. Traditionally, the conformal
fields $\varphi(x^\mu)$ posses a definite dilatation weight, just like the
Minkowski space coordinates $x^\mu$. This dilatation weight is the
representation label for the non-compact subgroup SO(1,1)$\sim \mathbb{R}$ of
the conformal group SO(4,2). However, the compact R symmetry group SO(6) does
not have such a subgroup, so the compensation can only take place between two
SO(2) subgroups of SO(4,2) and of SO(6). The group SO(2) is part of the maximal
compact subgroup ${\rm K} = {\rm SO}(4) \times{\rm SO}(2)$  of the conformal
group G = SO(4,2). Thus, it is natural to introduce conformal fields in such a
way that they form a {\it linear} representation of the maximal compact
subgroup K.

In general, the spaces in which the conformal fields are defined can be
regarded as coset spaces H$\backslash$G corresponding to different choices of
the subgroup H $\subset$ G. A formulation with a linear action of K should be
based on the coset space $\mathbb{H}^4_+$ = K$\setminus$G. This coset is a
four-dimensional non-compact K\"ahler manifold which can be viewed as a bounded
domain in $\mathbb{C}^4$. The free conformal fields correspond to the so-called
discrete series of UIRs of G. They can be realized as functions, analytic in
$\mathbb{H}^4_+$ and transforming homogeneously and irreducibly in
finite-dimensional UIRs of K. One says that such UIRs of G are ``induced" by
the compact subgroup K.

Apart from the conformal group, the fields of the ${\cal N}=4$ SYM theory are
also UIRs of the R symmetry group. It is then natural (and even necessary in
the contraction context, as we argue) to realize this internal symmetry in an
auxiliary compact space. It is obtained by dividing the R symmetry group SO(6)
by a maximal compact subgroup, which we choose to be the same ${\rm K} = {\rm
SO}(4) \times{\rm SO}(2)$. Only then, by examining the combined action of the
two SO(2) subgroups from the conformal and R symmetry sectors, we are able to
define a contraction limit in the Hilbert space of the conformal fields, such
that one of the two groups SO(2) survives. In this way we obtain the
Bargmann-Fock representation of the Heisenberg algebra, in which the generator
of the surviving SO(2) plays the r\^ole of the Hamiltonian. In the process of
contraction the bounded domain $\mathbb{H}^4_+$ is expanded to the full space
$\mathbb{C}^4$.

This picture is quite different from the traditional approach to conformal
fields in real (compactified) Minkowski space. There the UIRs of G are induced
by its parabolic subgroup consisting of the Lorentz group SO(1,3), dilatations
SO(1,1) and the Abelian subgroup of conformal boosts. The resulting space
$\overline{\mathbb{M}}^4$=(S$^1\times$ S$^3)/\mathbb{Z}_2$ can be viewed as the
boundary of the non-compact space $\mathbb{H}^4_+$. Consequently, the conformal
fields in $\overline{\mathbb{M}}^4$ are in fact the boundary values of the
analytic functions in $\mathbb{H}^4_+$. In particular, one can show that the
two spaces provide two different but equivalent realizations of the discrete
series of UIRs of G. Note that K acts as an isometry of
$\overline{\mathbb{M}}^4$ and one can label the conformal fields on
$\overline{\mathbb{M}}^4$ by the quantum numbers of K. The correlation
functions of the conformal fields on $\overline{\mathbb{M}}^4$ (the Wightman
functions) are also the limiting values of analytic functions on
$\mathbb{H}^4_+$.


Since in the contraction limit the bounded domain $\mathbb{H}^4_+$ blows up
into $\mathbb{C}^4$, it looses its boundary and, as a consequence, the
relationship with the representation theory in $\overline{\mathbb{M}}^4$
apparently breaks down. This clearly shows that the contraction, viewed by a
Minkowski space observer, is a singular process. At present we are not able to
exhibit a well-defined contraction limit for conformal fields initially defined
on $\overline{\mathbb{M}}^4$. This problem requires further study.

All the main features of the contraction procedure described above are already
seen in the simplest one-dimensional case. There the symmetry group is
SO(1,2)$\; \times\; $SO(3) $\sim$ SU(1,1)$\; \times\; $SU(2). The bounded
complex domain is the upper sheet of the double-sheeted hyperboloid
$\mathbb{H}^1_+ \sim$ SO(2)$\backslash$SO(1,2) (which is also isomorphic to the
Poincar\'{e} disc ${\cal D}^1$). The internal space is the two-sphere $S^2 \sim$
SO(2)$\backslash$SO(3). For the sake of simplicity, we restrict the discussion
in this paper to the one-dimensional case.

The paper is organized as follows. In Section 2 we briefly describe the
realizations of the bosonic algebra so(1,2)$\; \oplus\; $so(3) in different
coset spaces. In particular, we recall Dirac's manifestly covariant realization
of the conformal group on the light cone. It serves as a good introduction to
the approach to fields on coset spaces K$\backslash$G that we adopt: We regard
them as functions on the group G homogeneous under the left action of the
subgroup K. Section 3 contains some fairly standard material
\cite{Vilenkin,BarutFronsdal} about the UIRs of the groups SO(1,2) and SO(3).
In particular, we explain how the discrete series of UIRs of SO(1,2) (which
correspond to free conformal fields) can be constructed in a spinor basis. Then
we derive the matrix elements of these representations which are to be used as
an orthogonal basis for expanding conformal fields. Choosing to diagonalize
either the SO(2) or SO(1,1) subgroups of SO(1,2) leads to different notions of
conformal fields, carrying either a compact charge or a dilatation weight,
respectively. Only the former turns out suitable for the pp-wave limit. In
Section 4 we show how the free conformal fields living on the unit disc and on
the two-sphere can be contracted in the pp-wave limit. The main point here is
to expand the unit disc so that it covers the whole plane. In the process the
conformal weight and the R charge grow as the square of the disc radius. We
briefly discuss two possibilities to contract fields living in the compact
Minkowski space (the circle $S^1$ in this case) and explain why they fail.
Finally, in Section 5 we construct correlation functions of primary fields on
${\cal D}^1\times S^2$ and present a possible well-defined and non-trivial
contraction limit. In view of the holographic duality it would be interesting
to understand how these correlation functions could emerge from string theory.

\section{Generalities}

\subsection{The conformal and R symmetry algebras}

The bosonic part of the symmetry group of the ${\cal N}=4$ SYM theory in four
dimensions is SO(4,2)$\; \times\; $SO(6), which is locally isomorphic to
SU(2,2)$\; \times\; $SU(4). All the main features of the pp-wave limit that we
are discussing in this paper are perfectly well illustrated by the much simpler
one-dimensional case. So, we shall restrict ourselves to one-dimensional
conformal fields with symmetry group SO(1,2)$\; \times\; $SO(3) $\sim$
SU(1,1)$\; \times\; $SU(2).

The Lie algebra of SO(1,2) $\sim$ SU(1,1) $\sim$ SL(2,$\mathbb{R}$) with
antisymmetric generators $M_{ab}$, $M_{ab}=-M_{ba}$, has the following
commutation relations
\begin{eqnarray} \label{cr}
[M_{ab},M_{cd}]=\eta_{bc}M_{ad}+\eta_{ad}M_{bc}-\eta_{ac}M_{bd}-\eta_{bd}M_{ac}\,
, \end{eqnarray} where $a,b=0,1,2$ and the metric is $\eta={\rm diag }(-,+,-)$.
The non-compact group SO(1,2) has two important one-dimensional subgroups, the
non-compact SO(1,1) (generator $D\equiv M_{21}$) and the compact SO(2)
(generator $M_0 \equiv  iM_{02}$). It is useful to rewrite the algebra
(\ref{cr}) in two different bases with manifest covariance with respect to each
of these subgroups.

Introducing the notation $P=1/\sqrt{2}(M_{02}-M_{01})$ and
$K=1/\sqrt{2}(M_{01}+M_{02})$, we can recast (\ref{cr}) in the form
\begin{eqnarray}
  && [D,P]=-P\,, \nonumber\\
  && [D,K]=K\,, \label{crdil} \\
  && [P,K]=D\,. \nonumber
\end{eqnarray}
Here $D$, $P$ and $K$ are the familiar generators of dilatations, (time)
translations and conformal boosts, respectively.

The alternative basis is obtained by introducing complex combinations of
generators, $M_\pm = {1}/{\sqrt{2}}(iM_{12}\mp M_{01})$:
\begin{eqnarray}
  && [M_0,M_\pm]=\pm M_\pm\,,\nonumber\\
  && [M_+,M_-]=-M_0\,.\label{3}
\end{eqnarray}
Here the SO(2) $\sim$ U(1) generator $M_0$ and the raising $M_+$ and lowering
$M_-$ operators have the following properties under Hermitian conjugation:
\begin{equation}\label{2}
  M_0^\dagger = M_0\,, \qquad  M_+^\dagger = M_-\,.
\end{equation}

The representations of SO(1,2) are labeled by the quadratic Casimir
\begin{equation}\label{Casimir}
  C_2=D^2-PK-KP=M_0^2-M_+M_--M_-M_+ \, .
\end{equation}

The Lie algebra of SO(3) $\sim$ SU(2) with generators $L_{ab}$, $a,b=1,2,3$ has
the same commutation relations (\ref{cr}), except that the metric now is
$\eta_{ab}=\delta_{ab}$. This algebra admits a form analogous to (\ref{3}),
with the SO(2) $\sim$ U(1) generator $L_0= -iL_{12}$ and  the raising and
lowering operators $L_\pm = {i}/{\sqrt{2}}(L_{12}\pm i L_{13})$:
\begin{eqnarray}
  && [L_0,L_\pm]=\pm L_\pm\,,\nonumber\\
  && [L_+,L_-]=L_0\,,\label{1}
\end{eqnarray}
with the same properties (\ref{2}) under conjugation.   The only difference
between (\ref{1}) and (\ref{3}) is in the sign in front of the U(1) generator
$L_0$ or $M_0$. The expression for the quadratic Casimir becomes
\begin{equation}\label{Casimir'}
  C_2=L_0^2+L_+L_-+L_-L_+ \, .
\end{equation}

\subsection{Models of space-time}\label{Model}

Conformal fields can be introduced in several ways which are in correspondence
with different group-theoretic decompositions of an element of the conformal
group G. Below we list these possibilities for the one-dimensional case G =
SO(1,2).

\begin{figure}[ht]
\begin{center}
\input{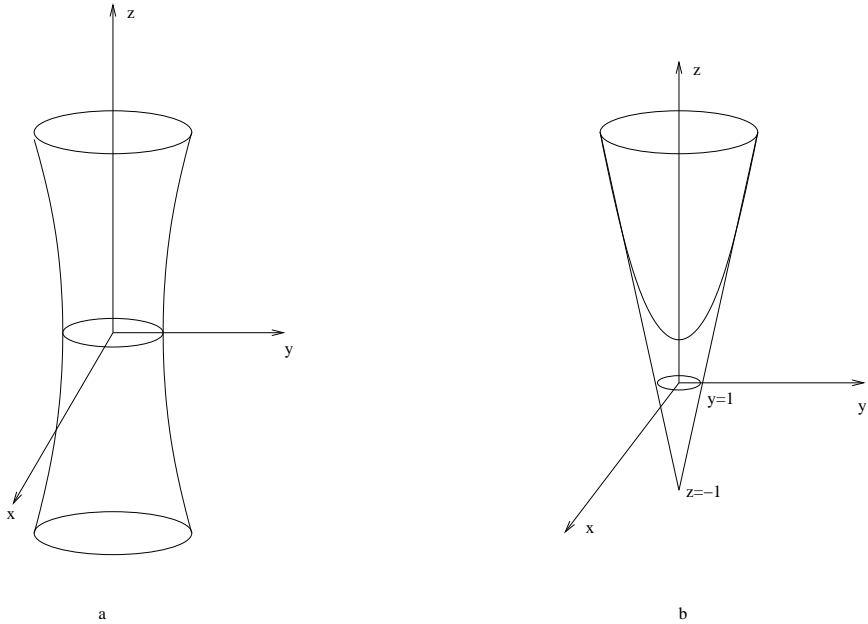}
\end{center}
\caption{Models of the coset spaces $\mathbb{H}_0= {\rm SO}(1,1)\backslash {\rm
SO}(1,2)$ and $\mathbb{H}_+= {\rm SO}(2)\backslash {\rm SO}(1,2)$. The
Poincar\'e disc ${\cal D}^1$ arises as the projection of the upper sheet of the
double-sheeted hyperboloid $\mathbb{H}_+$ on the $xy$-plane. Regarded as a
bounded complex domain, the disc is isomorphic to the coset space
U(1)$\backslash $SU(1,1).} \label{fig1}
\end{figure}

\subsubsection{The compact space} \label{sub1} The first way to define a
conformal field is based on the Bruhat decomposition G = NAM$\tilde{\mbox{N}}$
valid for almost all elements of the group.\footnote{For more details see
Section \ref{Mink}.} Here N is the nilpotent subgroup of special conformal
transformations (``boosts"), M is a center element (it becomes the Lorentz
group in higher dimensions), A = SO(1,1) is the group of dilatations and
$\tilde{\mbox{N}}$ the nilpotent subgroup of translations. The coset
space\footnote{The subgroup N is normal with respect to A, therefore the order
of the factors N and A can be changed. } NM$\backslash$G is the positive cone
C$_+$ in $\mathbb{R}^{1,2}$. Dividing further the cone by the dilatations we
obtain the conformal compactification of the Minkowski space:
$\overline{\mathbb{M}}^1 =S^1=$A$\backslash$C$_{+}$=ANM$\backslash\mbox{G}$.
Conformal fields are introduced as functions on C$_{+}$ transforming
homogeneously under A (see Section \ref{Dirac}). Thus, effectively they are
functions on the unit circle ANM$\backslash\mbox{G}=S^1$. \footnote{In physics
it is customary to replace the compact manifold $S^1$ by the one-dimensional
non-compact  Minkowski space $\mathbb{M}^1= \mathbb{R}^{1}$. The
compactification $\overline{\mathbb{M}}^1 = S^1$ is recovered by completing the
line $\mathbb{R}^{1}$ with the infinite point.} In this construction the
compact subgroup K = SO(2) plays the r\^ole of the isometry group of the
homogenous space $S^1$.

This coset space construction is depicted schematically in Figure 1a. The
non-compact coset SO(1,1)$\backslash $SO(1,2) is the single-sheeted
hyperboloid. The further factorization by the nilpotent subgroup N
corresponds to identifying the different sections (circles).

\subsubsection{The non-compact space} \label{sub2}
The second way is based on the Iwasawa decomposition G = KAN, where K is the
maximal compact subgroup. Dividing by K, we introduce the {\it non-compact}
coset space K$\backslash$G =AN. The conformal fields are described as functions
$f(g)$ on G obeying for any $k\in$K the covariance (homogeneity) condition
$f(kg)=\rho(k)f(g)$, where $\rho$ is a finite-dimensional unitary irreducible
representation (UIR) of K.

In the one-dimensional case K = SO(2) and the non-compact coset
SO(2)$\backslash$SO(1,2) is the upper sheet of the double-sheeted hyperboloid
(Figure 1b). It can be equivalently represented by its stereographic projection
onto the $xy$ plane, which is the open unit disc ${\cal D}^1$. The latter is
also the image of the isomorphic coset space U(1)$\backslash$SU(1,1).

The relation between the conformal spaces $S^1$ and  ${\cal D}^1$ introduced
above is established through the fact that the non-compact coset ${\cal D}^1$
is bounded and the boundary is precisely the compact Riemannian space $S^1$.
The non-compact coset is a K\"ahler manifold with boundary. Holomorphic
functions on this manifold are defined by their restriction to the boundary.
This is how the conformal fields on two different cosets are related to each
other.

It is important to realize that the above relationship between the compact and
non-compact pictures persists in the four-dimensional case as well. There the
compact Minkowski space is obtained by dividing the conformal group SO(4,2) by
its parabolic subgroup consisting of Lorentz transformations SO(3,1) (M),
dilatations SO(1,1) (A) and conformal boosts (N). The resulting space is
$\overline{\mathbb{M}}^4 = (S^1\times S^3)/\mathbb{Z}^2$. The non-compact space
is the coset K$\backslash $G = (SO(4)$\; \times\; $SO(2))$\backslash $SO(4,2)
$\sim$ S(U(2)$\; \times\; $U(2))$\backslash$SU(2,2). It is known \cite{Knapp}
that this space is a bounded complex domain whose boundary is precisely
$\overline{\mathbb{M}}^4$. Therefore we believe that our one-dimensional
considerations in this paper give a good idea of what happens in four
dimensions as well.

\subsubsection{The compact space of R symmetry}

The symmetry group we are discussing is not just the conformal group SO(1,2)
but its tensor product with the R symmetry group SO(3). Consequently, our
conformal fields are UIRs of both groups. While it is customary to realize the
UIRs of the conformal group in terms of fields, i.e., functions on a
space-time, in physics we usually exhibit the R symmetry by attaching indices
to the fields. However, for the purpose of studying the pp-wave contraction it
is necessary to treat both symmetry groups on the same footing. Thus, we are
lead to extend space-time by an auxiliary internal space, in which the R
symmetry group acts.\footnote{This idea is commonly used in the literature on
representation theory (see, e.g.,  \cite{Knapp,DMPPT}). As a side remark, note
that it is in the basis of the so-called ``harmonic superspace" approach to
extended supersymmetry, see \cite{Kniga,hh}.} Once again, it can be introduced
as a coset of the R symmetry group. In our simple case there is only one such
coset: the two-sphere $S^2\sim$ SO(2)$\backslash $SO(3) $\sim$ U(1)$\backslash
$SU(2), which is the simplest example of a compact complex manifold. Its
stereographic projection is the complex plane ${\mathbb C}^1$ with the infinite
point added.

In the four-dimensional case one has to divide the R symmetry group SO(6)
$\sim$ SU(4) by one of its maximal compact subgroups, K = SO(4)$\; \times\;
$SO(2) $\sim$ S(U(2)$\; \times\; $U(2)), which coincides with the maximal
compact subgroup of the conformal group SO(4,2). The resulting coset space is
again a compact complex manifold.

\subsection{The light cone as an example of a homogeneous space}\label{Dirac}

Here we recall Dirac's ``light ray" realization of  the coset space described
in \ref{sub1} (see e.g. \cite{MackSalam}).
It serves as a good introduction to the coset realizations we
use in Section \ref{u1}.

The starting point is the ``light cone" (the positive cone C$_+$ in
$\mathbb{R}^{1,2}$). It is described by a light-like vector $\xi^a$, $a=0,1,2$:
\begin{equation}\label{light1}
  \xi^a\xi_a = -(\xi^0)^2 +(\xi^1)^2 - (\xi^2)^2=0\,,
\end{equation}
where $\xi_a=\eta_{ab}\xi^b$. This condition is invariant under the linear
action of the conformal group SO(1,2). Its generators are given by
\begin{equation}\label{light2}
  M_{ab} = \xi_a \pa/\pa\xi^b - \xi_b \pa/\pa\xi^a\,
\end{equation}
and satisfy the algebra (\ref{cr}).

The conformal algebra can be realized on fields $\Phi(\xi)$ on the light cone,
homogeneous of degree $\ell$, $\Phi(\rho\xi) = \rho^\ell\Phi(\xi)$. This
condition effectively eliminates one degree of freedom, so we are left with a
single independent variable. In this way we introduce the one-dimensional
Minkowski coordinate (time) $x\equiv x^0$ as a {\it projective coordinate}:
\begin{equation}\label{light3}
  x = \frac{\xi^0}{\xi^1+\xi^2}\,.
\end{equation}
We have chosen to divide by the combination $\xi^1+\xi^2$ because it transforms
homogeneously under dilatations, as can easily be verified. Thus, the new
coordinate $x$ has a definite dilatation weight.

The Minkowski fields $\varphi(x)$ are then introduced through the relation
\begin{equation}\label{light4}
  \Phi(\xi)  = (\xi^1+\xi^2)^{-\Delta}\varphi(x)\,,
\end{equation}
where $\Delta=-\ell>0$ is the conformal dimension of the field. In this form it
is obvious that $\Phi(\xi)$ is homogeneous of degree $\ell$, since the
coordinate $x$ is inert under the rescaling $\xi^a\rightarrow \rho\xi^a$.

The principal difference between the fields $\Phi(\xi)$ and $\varphi(x)$ is
that the former is manifestly covariant under the conformal group SO(1,2),
while the latter transforms non-linearly. To see this, it is instructive to
rewrite the generators (\ref{light2}) in terms of the single variable $x$. With
the help of (\ref{light1})-(\ref{light4}) it is not hard to derive the familiar
form of the generators of the conformal algebra  (\ref{crdil}):
\begin{eqnarray}
  &&D = x\frac{d}{dx} + \Delta \nonumber\\
  &&P =\frac{1}{\sqrt{2}}\frac{d}{dx}\label{light5}\\
  &&K =  \frac{1}{\sqrt{2}}\left(x^2 \frac{d}{dx} + 2\Delta x\right)  \nonumber
\end{eqnarray}
It is well known that such fields realize irreducible representations of the
conformal group of weight $\Delta$. Indeed, on $\varphi(x)$ the Casimir
(\ref{Casimir}) takes the value $C_2=\Delta(\Delta-1)$.

The construction above provides a realization of the coset space
$\mathbb{R}^{1}\backslash$C$_+$. We start with functions defined on the bigger
space C$_+$ and impose a homogeneity condition with respect to the  ``stability
group" $\mathbb{R}^{1}\sim$ SO(1,1). The degree of homogeneity corresponds to
the weight, i.e., to the representation label. The resulting fields depend on
the ``carrier" coordinate (in eq. (\ref{light4}) it is $\xi^1+\xi^2$) for the
stability group through a factor determined by the weight. In what follows we
shall apply this coset construction to the homogeneous spaces K$\backslash$G,
where K = U(1) and G = SU(2) or SU(1,1). Our fields will be defined on the
whole group G, but they will be homogeneous under the action of the subgroup K.
The manifest covariance under K will be achieved at the expense of introducing
an extra coordinate for it. However, the dependence on this ``weight carrier"
coordinate will always be factorized.

\section{Free conformal fields}

In this section we describe the lowest-weight UIRs of the conformal and R
symmetry group using a spinor basis. Next we calculate the matrix elements of
the representations and show that a particular subset of them can be used as an
orthonormal basis for expanding free conformal fields of a given weight. Then
we introduce different coordinatizations by decomposing the group elements
according to the patterns discussed in Section \ref{Model}.

\subsection{Lowest-weight unitary representations of SU(2) and SU(1,1)}

We start by recalling some basic facts\footnote{Our presentation follows Ref.
\cite{BarutFronsdal}.} about the representations with lowest weight, commonly
used in physics, both in the familiar case of SU(2) as well as the more
elaborate case of SU(1,1). The representations of these two groups (and, more
generally, of their common complex form SL(2,$\mathbb{C}$)) are most
conveniently realized \cite{Waerden1932,Wigner1959,BarutFronsdal} in a vector
space spanned by complex monomials:
\begin{equation}\label{4}
  |a,b\rangle = z_1^az_2^b\,.
\end{equation}

It is well known that the UIRs of the compact group SU(2) are
finite-dimensional and are given by the homogeneous polynomials of degree
$2\ell$ formed by the monomials:
\begin{equation}\label{5}
  \mbox{SU(2)}: \qquad |\ell,m\rangle = N_m z_1^mz_2^{2\ell-m}\,, \qquad \ell\geq0,
  \quad m=0,1,\ldots,2\ell\,.
\end{equation}
In order to determine the normalization coefficients $N_m$, let us first write
down the generators (\ref{1}) explicitly:
\begin{equation}\label{6}
  L_0 = \frac{1}{2}\left(z_1{\pa}/{\pa z_1} - z_2{\pa}/{\pa
  z_2}\right)\,, \quad L_+ = \frac{1}{\sqrt{2}}\, z_1{\pa}/{\pa z_2} \,,
  \quad L_- = \frac{1}{\sqrt{2}}\, z_2{\pa}/{\pa
  z_1} \,.
\end{equation}
Their action on the basis vectors is given by
\begin{eqnarray}
  &&L_0|\ell,m\rangle = (m-\ell)|\ell,m\rangle \nonumber\\
  &&L_+|\ell,m\rangle = \frac{2\ell-m}{\sqrt{2}} \frac{N_m}{N_{m+1}}|\ell,m+1\rangle \label{7} \\
  &&L_-|\ell,m\rangle = \frac{m}{\sqrt{2}} \frac{N_m}{N_{m-1}}|\ell,m-1\rangle
    \nonumber
\end{eqnarray}
Note that $|\ell,m\rangle$ are eigenvectors of the U(1) generator $L_0$ with
eigenvalue (``charge") $m-\ell$.

Assuming that the basis is orthonormal,
\begin{equation}\label{8}
 \langle \ell, m|\ell,n\rangle = \delta_{mn}\,,
\end{equation}
we can translate the unitarity condition (\ref{2}) into a condition on the
normalization coefficients $N_m$:
\begin{equation}\label{9}
  \left\vert\frac{N_{m+1}}{N_m}\right\vert^2 = \frac{2\ell-m}{m+1} \geq 0\,.
\end{equation}
It has the following solution
\begin{equation}\label{10}
  N_m = \left(\frac{(2\ell)!}{{m!(2\ell-m)!}}\right)^{1/2}\,,
\end{equation}
where we have set $N_0=1$.  We remark that the range of the index $m$ shown in
(\ref{5}) is in fact a consequence of unitarity, as is clear from eq.
(\ref{9}).

The case of the non-compact group SU(1,1) is considerably more involved. This
group is known to possess four series of infinite-dimensional UIRs
\cite{Bargmann, Vilenkin}: principal, supplementary and two discrete series
${\cal D}^\pm_\ell$. All of them admit a realization on vectors of the type
(\ref{1}). However, only the discrete series are suitable for applications in
CFT, since only they have a lowest weight vector (for ${\cal D}^+_\ell$) or a
highest weight vector (for ${\cal D}^-_\ell$), and, correspondingly, their
energy spectrum is bounded below or above \cite{Mack}. Here we restrict
ourselves to the discrete series ${\cal D}^+_\ell$.

The first modification, compared to the case SU(2), is in the expression for
the SU(1,1) generators (\ref{3}) (cf. (\ref{6})):
\begin{equation}\label{11}
  M_0 = \frac{1}{2}\left(z_1\pa/{\pa z_1} - z_2\pa/{\pa
  z_2}\right)\,, \quad M_+ = -\frac{i}{\sqrt{2}} \, z_1\pa/{\pa z_2} \,,
  \quad M_- = -\frac{i}{\sqrt{2}}\, z_2\pa/{\pa
  z_1} \,.
\end{equation}
We use the same basis (\ref{5}), but now keeping $m$ and $\ell$ arbitrary (in
general, complex). After making the corresponding changes in (\ref{7}), we see
that if neither $m$ nor $2\ell-m$ are integers, there can be no lowest (or
highest) weight vector. This case corresponds to the principal or supplementary
series of representations. The vector with $m=0$ is the lowest weight vector of
the invariant subspace spanned by $m=0,1,2,\ldots$. This case corresponds to
the discrete series ${\cal D}^+_\ell$.

Further, we have to examine the condition for unitarity, which now reads
\begin{equation}\label{12}
  \left\vert\frac{N_{m+1}}{N_m}\right\vert^2 = \frac{m-2\ell}{m+1} \geq 0\,.
\end{equation}
In the case of the discrete series ${\cal D}^+_\ell$  $m=0,1,2,\ldots$,
therefore we have to take a strictly negative weight $\ell<0$ (we exclude the
trivial representation $m=\ell=0$). In addition, for our illustrative purposes
it will be sufficient to restrict ourselves to integer or half-integer values
of $\ell$ only.\footnote{We remark that the name ``discrete series" comes from
precisely this restriction, which is common in the mathematical literature
\cite{Bargmann,Vilenkin}. Allowing for non-integer weight $2\ell$ (or
``anomalous dimension" in the physics terminology) leads to mathematical
subtleties which we need not discuss here. If one insists on the
representations of the algebra being integrable to representations of the
group, one is confronted with infinitely multivalued functions
\cite{FradkinPalchik}. Another way to see the problem is to switch from the
group SU(1,1) to its infinitely sheeted universal covering \cite{Mack,Pisa}.
For a discussion from the physicist's point of view see also Ref.
\cite{Witten}. } Thus, the vector space of the representation ${\cal D}^+_\ell$
becomes (cf. (\ref{5}))
\begin{equation}\label{13}
  \mbox{SU(1,1)}: \qquad |\ell,m\rangle = N_m z_1^mz_2^{2\ell-m}\,, \qquad \ell<0,
  \quad m=0,1,\ldots,\infty\,,
\end{equation}
where
\begin{equation}\label{14}
  N_m = \left(\frac{(m-2\ell-1)!}{{m!(-2\ell-1)!}}\right)^{1/2}\,.
\end{equation}

We note an important difference between the finite-dimensional UIRs of SU(2)
and the infinite-dimensional lowest weight UIRs of SU(1,1). The eigenvalue of
the U(1) generator $L_0$ or $M_0$ on the lowest weight vector  (``vacuum")
$|\ell,0\rangle$, which is $-\ell$ in both cases, is {\it negative} for SU(2)
and  {\it positive} for SU(1,1). So, this eigenvalue, which can be related to
the physical quantity ``energy" \cite{Mack}, stays positive within the
infinite-dimensional space of an SU(1,1) UIR of the class ${\cal D}^+_\ell$. It
is customary to label the UIRs of both groups by a positive number, the
``isospin" $J=\ell$ for SU(2) and the ``conformal weight" $\Delta=-\ell$ for
SU(1,1). It then easily follows that the quadratic Casimirs (\ref{Casimir'}) of
SU(2) and (\ref{Casimir}) of SU(1,1) take the values $C_2^{SU(2)} = J(J+1)$ and
$C_2^{SU(1,1)} = \Delta(\Delta-1)$.

\subsection{Matrix elements of SU(2) and SU(1,1)}

The matrix elements of SU(2) or SU(1,1) in the bases (\ref{5}) or (\ref{13})
can be obtained by considering the pair of complex numbers $z_1,z_2$ as the
fundamental spinor representation\footnote{The group SL(2,$\mathbb{C}$) has a
second fundamental representation obtained by complex conjugation.} of the
complexified group SL(2,$\mathbb{C}$). For any SL(2,$\mathbb{C}$) matrix
\begin{equation}\label{15}
  g=\left(
  \begin{array}{ll}
    \alpha & \beta \\
    \gamma & \delta
  \end{array}
  \right), \qquad \alpha\delta - \beta\gamma = 1
\end{equation}
we have
\begin{equation}\label{16}
  T(g) \left(
  \begin{array}{l}
   z_1 \\
    z_2
  \end{array}
  \right) =   \left(
  \begin{array}{ll}
    \alpha z_1 + \beta z_2\\
    \gamma z_1 + \delta z_2
  \end{array}
  \right) \,.
\end{equation}
Let us apply this rule to the basis vectors $|\ell,m\rangle$ and expand the
result in the same basis:
\begin{equation}\label{17}
  T^\ell(g)|\ell,m\rangle  = N_m (\alpha z_1 + \beta z_2)^m
  (\gamma z_1 + \delta z_2)^{2\ell-m} = \sum_n t^\ell_{mn}(g)N_n z^n_1
  z_2^{2\ell-n}\,.
\end{equation}
Here the sum over $n$ runs from 0 to either $2\ell$ (for SU(2)) or $\infty$
(for SU(1,1)). The matrices $t^\ell_{mn}(g)$ are the matrix elements of the
representation of weight $\ell$ corresponding to the group element $g$
(\ref{15}). Later on we shall specify $g$ to an SU(2) or an SU(1,1) matrix
written down in different parametrizations, but for the moment we keep the
SL(2,$\mathbb{C}$) notation.

Now, consider the subgroup U(1) $\times $SO(1,1) $ \subset$ SL(2,$\mathbb{C}$),
i.e., the group of diagonal matrices
\begin{equation}\label{15'}
  h=\left(
  \begin{array}{ll}
    \alpha^{1/2} & 0 \\
    0 & \alpha^{-1/2}
  \end{array}
  \right)\,.
\end{equation}
{}From (\ref{17}) we have
\begin{equation}\label{16'}
  t^\ell_{mn}(h) = \delta_{mn} \alpha^{m-\ell}\,.
\end{equation}
Thus, using the group multiplication property, we obtain the transformation law
of the matrix elements $t^\ell_{mn}(g)$ under the left action of the subgroup
(\ref{15'}):
\begin{equation}\label{17'}
  t^\ell_{mn}(hg) = \sum_k t^\ell_{mk}(h) t^\ell_{kn}(g) =
  \alpha^{m-\ell} t^\ell_{mn}(g)\,.
\end{equation}
In particular, setting $m=0$, we find
\begin{equation}\label{18'}
  t^\ell_{0n}(hg) = \alpha^{-\ell} t^\ell_{0n}(g)\,.
\end{equation}
The conclusion is that the matrix elements $t^\ell_{0n}(g)$, which correspond
to the lowest weight vector, transform homogeneously with weight $-\ell$.

Another important property of the matrix elements $t^\ell_{mn}(g)$ is their
orthogonality with respect to a suitable scalar product in the group space.
Moreover, under certain conditions they form complete bases for expanding whole
classes of functions on the group or on its factor spaces. A well-known example
is provided by the spherical harmonic functions for SU(2). In our notation they
correspond to the matrix elements
\begin{equation}\label{SphHar}
  Y_{\ell,\ell-n} \ \leftrightarrow \ t^\ell_{\ell n}\,.
\end{equation}
They are used for expanding (square integrable) functions on the sphere $S^2$,
{\it invariant} under the action of SU(2) (``scalar fields"). Indeed, from
(\ref{17'}) it is clear that $t^\ell_{\ell n}$ have vanishing weight. Note that
an expansion in terms of the spherical harmonics constitutes an infinitely
reducible representation of SU(2). What we need here is different.

As explained at the end of Section \ref{Dirac}, for us the conformal fields are
defined as functions on the group G  {\it transforming homogeneously} under the
left action of the subgroup K and being  {\it irreducible} under both G and K.
This gives an effective realization of the coset K$\backslash $G. In this paper
we consider the cosets U(1)$\backslash$SU(2) and U(1)$\backslash$SU(1,1). In
both cases K is diagonal, so the discussion above applies. So, we deduce that
such functions can be expanded in terms of the matrix elements $t^\ell_{0n}(g)$
corresponding to the lowest weight vector of the representation $T^\ell(g)$:
\begin{equation}\label{251}
  \Phi^\ell(g) =  \sum_{n} a_n t^\ell_{0n}(g)\, .
\end{equation}
Under the diagonal subgroup $h$  (\ref{15'}) they transform homogeneously with
weight $-\ell$:
\begin{equation}\label{252}
  \Phi^\ell(hg) = \alpha^{-\ell}\Phi^\ell(g)\,.
\end{equation}
At the same time, the {\it positive} number $|\ell|$ labels the UIR of the
group $g$ realized on the function $\Phi^\ell$.

The matrix elements $t^\ell_{0n}(g)$ are particularly easy to
derive.
Indeed, setting $m=0$ in (\ref{17}) we obtain
\begin{equation}\label{18}
   (\gamma z_1 + \delta z_2)^{2\ell} = \sum_n t^\ell_{0n}(g)N_n z^n_1
  z_2^{2\ell-n}\,.
\end{equation}
The expansion of the left-hand side is performed differently for SU(2) and
SU(1,1). In the case of SU(2) we have $\ell\geq 0$, so this is just a
binomial expansion, valid for any values of $\gamma, \delta$ and $z_{1,2}$.
After equating the coefficients of the terms $z^n_1
  z_2^{2\ell-n}$, we find
\begin{equation}\label{19}
 \mbox{SU(2)}: \qquad  t^\ell_{0n}(g) = \left(\frac{(2\ell)!}
 {n!{(2\ell-n)!}}\right)^{1/2}\, \gamma^n \delta^{2\ell-n}\,.
\end{equation}

In the case of SU(1,1) we have  $\ell< 0$, so we need to define the parameter
domain in which we can expand. For instance, we can set
\begin{equation}\label{20}
  |\gamma|<|\delta|\,, \qquad |z_1|<|z_2|
\end{equation}
and rewrite the left-hand side of eq. (\ref{18}) as follows:
\begin{equation}\label{21}
   (\gamma z_1 + \delta z_2)^{2\ell} = (\delta z_2)^{2\ell} \left(1 +
   \frac{\gamma}{\delta} \frac{z_1}{z_2}\right)^{-2|\ell|}\,,
\end{equation}
and then expand. As a result, we obtain the matrix elements
\begin{equation}\label{22}
\mbox{SU(1,1)}: \qquad  t^\ell_{0n}(g) = (-1)^n\left(\frac{(n-2\ell-1)!}
 {n!{(-2\ell-1)!}}\right)^{1/2}\, \gamma^n \delta^{2\ell-n}\,.
\end{equation}

\subsection{Coordinates on the homogeneous spaces U(1)$\backslash$SU(2) and U(1)$\backslash$SU(1,1)
}\label{u1}

As explained above, our conformal fields are defined on coset spaces of the
conformal group SU(1,1) and the $R$ symmetry group SU(2). The matrix
elements introduced in the preceding subsection provide a natural basis for
expanding such functions.

\subsubsection{The compact case} \label{leftrightsect}

Consider first the left coset U(1)$\backslash$SU(2)$\sim S^2$. We can introduce
coordinates on it by specifying the matrix (\ref{15}) to SU(2) and writing it
down in a factorized form, where the U(1) factor stands on the left:
\begin{equation}\label{23}
  g=\left(
  \begin{array}{cc}
    e^{i\phi/2} & 0 \\
    0 & e^{-i\phi/2}
  \end{array}
  \right)\, \frac{1}{\sqrt{1+|\tau|^2}}\, \left(
  \begin{array}{rr}
    1 & -\bar\tau \\
    \tau & 1
  \end{array}
  \right)\,.
\end{equation}
Here $-2\pi \leq \phi < 2\pi$ is the coordinate of the subgroup U(1) and $\tau$
is the complex stereographic coordinate on the coset U(1)$\backslash$SU(2)$\sim
S^2$. Reading $\gamma$ and $\delta$ off from (\ref{23}) and substituting in
(\ref{19}), we obtain
\begin{equation}\label{24}
  t^\ell_{0n} = \left(\frac{(2\ell)!}
 {n!{(2\ell-n)!}}\right)^{1/2}\, e^{-i\ell\phi} \frac{\tau^n}{(1+|\tau|^2)^\ell}\,.
\end{equation}
According to our definition (\ref{251}), a ``field", i.e. a function carrying
the irreducible representation of lowest weight $-\ell$, has the form
\begin{equation}\label{25}
  \Phi^\ell(\tau ,\bar\tau,\phi) =
  \sum_{n=0}^{2\ell}a_n t^\ell_{0n}(\tau ,\bar\tau,\phi) =
   \frac{e^{-i\ell\phi}}{(1+|\tau|^2)^\ell}
  \sum_{n=0}^{2\ell}b_n \tau^n \equiv e^{-i\ell\phi}\varphi(\tau ,\bar\tau)\,.
\end{equation}
We clearly see that the dependence on the U(1) coordinate $\phi$ is given by an
overall weight factor, which reflects the fact that the field $\Phi^\ell$ is
effectively defined on the coset U(1)$\backslash$SU(2)$\sim S^2$. In a way, the
picture is very similar to Dirac's light-cone description of conformal fields:
Eq. (\ref{25}) is the analog of eq. (\ref{light4}). The important difference,
however, is that the factor in the right-hand side of eq. (\ref{25}) comes from
the compact subgroup SO(2), while that in eq. (\ref{light4}) is related to the
non-compact SO(1,1).

Further, the dependence on the complex coordinates $\tau,\bar\tau$ in
(\ref{25}) is essentially holomorphic, i.e. it is determined by a polynomial
$\sum_{n=0}^{2\ell}b_n \tau^n$ of degree $2\ell$ . The non-holomorphic factor
$1/(1+|\tau|^2)^\ell$ can be combined with the SU(2) invariant measure. The
scalar product in the space of the functions (\ref{25}) is given by
\begin{eqnarray}
 \langle \Phi^\ell_1 \Phi^{\ell'}_2 \rangle &=& \frac{\sqrt{(2\ell+1)(2\ell'+1)}}{4\pi^2}
  \int_{\mathbb{C}}\frac{d\tau d\bar\tau }{(1+|\tau|^2)^2}
   \int^{2\pi}_{-2\pi} d\phi\ \Phi^\ell_1 \bar\Phi^{\ell'}_2 \nonumber\\
  &=&\frac{2\ell+1} {\pi} \delta^{\ell\ell'} \int_{\mathbb{C}}\frac{d\tau d\bar\tau }{(1+|\tau|^2)^2}
  \ \varphi_1(\tau,\bar\tau) \Bar\varphi_2(\tau,\bar\tau) \,. \label{25'}
\end{eqnarray}
It is easy to check that the matrix elements (\ref{24}) form an orthonormal set
with respect to this scalar product:
\begin{equation}\label{25''}
  \langle t^\ell_{0n} t^{\ell'}_{0n'}\rangle = \delta^{\ell\ell'}
  \delta_{nn'}\,.
\end{equation}

Let us address the question of how the SU(2) algebra is realized on such
functions. By construction, they are invariant under the right action of G. To
put it differently, the right action on the coordinates can be compensated by a
suitable transformation of the coefficients $b_n$ in (\ref{25}), which form a
UIR of G on their own.\footnote{An analogy is provided by a four-dimensional
scalar field written down as a power expansion, $\phi(x^\mu) = f + f_\mu x^\mu
+ f_{\mu\nu} x^\mu x^\nu + \ldots $. The coefficients in this expansion $f,
f_\mu, f_{\mu\nu},\ldots $ are a Lorentz scalar, vector, tensor, etc., although
the field $\phi(x^\mu)$ itself is a Lorentz scalar.} Then it is natural to
realize the SU(2) algebra in terms of the left-covariant Lie derivatives on the
coset K$\backslash$G. Their explicit form can be derived using the standard
Cartan method of differential forms. It consists in expanding the form
$g^{-1}dg$ in the basis of the Lie algebra generators
\begin{equation}\label{cartan}
  L_0 = -\frac{1}{2}\left(
  \begin{array}{rr}
    1 & 0 \\
    0 & -1
  \end{array}
  \right)\,, \qquad   L_+ = \frac{1}{\sqrt{2}}\left(
  \begin{array}{rr}
    0 & 0 \\
    1 & 0
  \end{array}
  \right)\,, \qquad  L_- = \frac{1}{\sqrt{2}}\left(
  \begin{array}{rr}
    0 & 1 \\
    0 & 0
  \end{array}
  \right)
\end{equation}
and then inverting the coefficient matrix. An alternative way (the simplest) is
to directly compute the left action of the group on the matrices (\ref{23}) and
extract from it the action of the generators (\ref{cartan}) on the coordinates.
Yet another way \cite{Kniga}, which is suitable for making contact with the
discussion in Section \ref{Dirac}, is to realize the generators in terms of the
entries $g_\pm{}^i$ of an abstract SU(2) matrix.  Here $\pm$ refers to the
first or second row of the matrix, according to the sign in $e^{\pm i\psi/2}$.
They satisfy the unit determinant condition
\begin{equation}\label{unitdet}
  g_+{}^i\epsilon_{ij} g_-{}^j = 1\,.
\end{equation}
Now, let us write down the following differential operators:
\begin{eqnarray}
  && L_0 = g_+{}^i {\partial\over \partial g_+{}^i} -
  g_-{}^i {\partial\over \partial g_-{}^i}\,, \nonumber\\
  && L_- = g_-{}^i {\partial\over \partial g_+{}^i}\,, \qquad
  L_+ = g_+{}^i {\partial\over \partial g_-{}^i}\,.   \label{harmder}
\end{eqnarray}
Here we can formally treat the variables $g_\pm{}^i$ as independent, since all
the three operators are compatible with the condition (\ref{unitdet}). Then it
is almost obvious that they form the SU(2) algebra  (\ref{1}). By construction,
these generators are right-invariant (the right-hand side indices $i$ of
$g_\pm{}^i$ are contracted), but they have charges under the left group. It is
then a straightforward exercise to substitute the explicit coordinatization
(\ref{23}) and obtain the left covariant derivatives:
\begin{eqnarray}
  && L_0 = -i {\partial\over \partial\phi} \nonumber\\
  && L_- = \frac{e^{-i\phi}}{\sqrt{2}}
  \left[i(1+|\tau|^2){\partial\over \partial \bar\tau} -
{\tau } {\partial\over \partial\phi}\right] \nonumber\\
  && L_+ = \frac{e^{i\phi}}{\sqrt{2}} \left[i(1+|\tau|^2){\partial\over \partial\tau}
  + {\bar\tau } {\partial\over \partial\phi}\right]\,. \label{2500}
\end{eqnarray}
Applying the first two of these generators to the field $\Phi^\ell$ (\ref{25}),
we see that it can be regarded as a lowest weight vector of the left algebra
(\ref{2500}), in accord with our definition of the field:
\begin{equation}\label{leftlws}
  L_0 \Phi^\ell = -\ell \Phi^\ell\,, \qquad L_- \Phi^\ell = 0\,.
\end{equation}

We stress once more that our fields $\Phi^\ell(\tau ,\bar\tau,\phi)$ (\ref{25})
are defined as homogeneous functions on the whole of SU(2). If we want to deal
with functions on the sphere $S^2\sim$U(1)$\backslash$SU(2) itself, we need a
different realization of the SU(2) algebra, this time corresponding to the
right action of the group. The situation here is similar to  Section
\ref{Dirac}. There we started with a manifestly covariant form of the SO(1,2)
generators, eq. (\ref{light2}). In fact, the vector $\xi^a$ can be regarded as
a combination of two rows of an SO(1,2) matrix $g_A{}^a$ having definite
SO(1,1) weight under the left action of the group, $\xi^a = g_0{}^a + g_2{}^a$
(the light-cone condition (\ref{light1}) then follows from the
pseudo-orthogonality of this matrix). The operators $M_{ab}$ are constructed so
that they have vanishing left weight, but are manifestly covariant under the
right action. By analogy, in the case of SU(2) we introduce the generators
\begin{equation}\label{rightgen}
  L^i{}_j = g_+{}^i {\partial\over \partial g_+{}^j}  - g_-{}^i {\partial\over \partial
  g_-{}^j} - \mbox{trace in $i,j$}\,.
\end{equation}
Once again, they form the algebra of SU(2), but this time written in spinor
notation. Like (\ref{light2}), these generators are left-invariant, but
right-covariant. Substituting the explicit coordinatization (\ref{23}),
changing from spinor to vector notation and applying the generators to the
function $\varphi(\tau ,\bar\tau)$ in (\ref{25})  (cf. (\ref{light4})), we
obtain the right algebra realized on functions on the sphere $S^2$:
\begin{eqnarray}
  && \widetilde L_0 = \tau{\partial\over \partial \tau} -
  \bar \tau{\partial\over \partial \bar \tau} - \ell \nonumber\\
  && \widetilde L_+ = \frac{i}{\sqrt{2}}
  \left[{\partial\over \partial \bar\tau} + \tau^2 {\partial\over \partial
  \tau} -\ell\tau \right] \nonumber\\
  && \widetilde L_- = \frac{i}{\sqrt{2}}
  \left[{\partial\over \partial \tau} + \bar\tau^2 {\partial\over \partial
  \bar\tau} + \ell\bar\tau \right]\,.  \label{2500'}
\end{eqnarray}


\subsubsection{The non-compact case}

The case of SU(1,1) can be treated in the same way. We consider the left coset
U(1)$\backslash$SU(1,1)$\sim {\cal D}^1 \sim \mathbb{H}_+$ (see Section
\ref{sub2}). Now the factorization (\ref{23}) is replaced by
\begin{equation}\label{26}
  g=\left(
  \begin{array}{cc}
    e^{i\psi/2} & 0 \\
    0 & e^{-i\psi/2}
  \end{array}
  \right)\, \frac{1}{\sqrt{1-|t|^2}}\, \left(
  \begin{array}{rr}
    1 & \bar t \\
    t & 1
  \end{array}
  \right)\,, \qquad |t|<1\,,
\end{equation}
where the complex variable $t$ describes the interior of the unit disc. The
analogs of eqs. (\ref{24}), (\ref{25}) and (\ref{25'}) are
\begin{equation}\label{27}
  t^\ell_{0n} = (-1)^n\left(\frac{(n-2\ell-1)!}
 {n!{(-2\ell-1)!}}\right)^{1/2}\, e^{-i\ell\psi} \frac{t^n}{(1-|t|^2)^\ell}\,,
\end{equation}
\begin{equation}\label{28}
  \Phi^\ell(t,\bar t,\psi) =  \sum_{n=0}^{\infty}a_n t^\ell_{0n}(t,\bar t,\psi) =
   \frac{e^{-i\ell\psi}}{(1-|t|^2)^\ell}
  \sum_{n=0}^{\infty}b_n t^n \equiv e^{-i\ell\psi}\varphi(t,\bar t)\,
\end{equation}
and
\begin{equation}\label{28'}
  \langle \Phi^\ell_1 \Phi^{\ell'}_2 \rangle = \frac{\sqrt{(2\ell+1)(2\ell'+1)}}{4\pi^2}
  \int_{|t|<1}\frac{dtd\bar t }{(1-|t|^2)^2} \int^{2\pi}_{-2\pi} d\psi\ \Phi^\ell_1 \bar\Phi^{\ell'}_2 \, ,
\end{equation}
where the positive value of the square root is assumed in the r.h.s. This time
the field is determined by the function $\sum_{n=0}^{\infty}b_n t^n$,
holomorphic inside the unit disc, and the factor $1/(1-|t|^2)^\ell$ can be
combined with the SU(1,1) invariant measure.   The left-covariant derivatives
on the coset U(1)$\backslash$SU(1,1) are given by (cf. (\ref{2500}))
\begin{eqnarray}
  && M_0 = -i {\partial\over \partial\psi} \nonumber\\
  && M_- = \frac{e^{-i\psi}}{\sqrt{2}} \left[i(1-|t|^2){\partial\over \partial\bar t}
  +{t } {\partial\over \partial\psi}\right] \nonumber\\
  && M_+ = \frac{e^{i\psi}}{\sqrt{2}} \left[i(1-|t|^2){\partial\over \partial t}
  - {\bar t } {\partial\over \partial\psi}\right]\,. \label{2501}
\end{eqnarray}
As before (cf. (\ref{leftlws})), the fields $\Phi^\ell$ are lowest weight
vectors for the left algebra (\ref{2501}):
\begin{equation}\label{leftlws'}
  M_0 \Phi^\ell = -\ell \Phi^\ell\,, \qquad M_- \Phi^\ell = 0\,.
\end{equation}

We can switch from homogeneous functions on the whole of SU(1,1) to functions
$\varphi(t,\bar t)$ (\ref{28})
on the disc ${\cal D}^1$ by considering the right action of SU(1,1).
The generators of the right algebra realized on functions on the disc
are
\begin{eqnarray}
  && \widetilde M_0 = t{\partial\over \partial t} -
  \bar t {\partial\over \partial \bar t} - \ell \nonumber\\
  && \widetilde M_+ = \frac{i}{\sqrt{2}}
  \left[{\partial\over \partial \bar t} - t^2 {\partial\over \partial
  t} +\ell t \right] \nonumber\\
  && \widetilde M_- = \frac{i}{\sqrt{2}}
  \left[{\partial\over \partial t} - \bar t^2 {\partial\over \partial
  \bar t} - \ell\bar t \right]\,. \label{discgen}
\end{eqnarray}

\subsection{Minkowski space as a coset space}\label{Mink}

The realization of the conformal group in terms of holomorphic functions in the
complex coset space U(1)$\backslash$SU(1,1) is very convenient for studying the
UIRs. However, in physics we are accustomed to a different realization, in {\it
real Minkowski space}.  A characteristic feature of this description is that
the fields $\phi(x)$ transform homogeneously under the non-compact dilatation
subgroup SO(1,1)$ \subset$ SU(1,1), and not under the compact SO(2) $ \subset$
SU(1,1), as was the case in the preceding subsection. There are many papers and
books where this subject is treated in great detail (see, e.g.,
\cite{MackSalam,DMPPT,Pisa}). Here we want to present a somewhat different picture,
along the lines of Section \ref{u1}. Our aim is to find a set of basis vectors
for Minkowski fields, manifestly covariant under SO(1,1). This will be used in
the discussion of the contraction mechanism in Section \ref{contr}. Our
presentation is rather heuristic, a rigorous treatment can be found in
\cite{Vilenkin}.

The group SU(1,1) admits several decompositions. The Cartan decomposition KAK,
where K = SO(2) and A = SO(1,1) is the factorization (\ref{26}) of the SU(1,1)
matrix. The Iwasawa decomposition G = KAN is
\begin{equation}\label{29'}
  g=
   \left(
  \begin{array}{cc}
    e^{i\theta/2} & 0 \\
    0 & e^{-i\theta/2}
  \end{array}
  \right)\,
\left(
  \begin{array}{cc}
    \cosh\frac{a}{2} & i\sinh\frac{a}{2} \\
   -i\sinh\frac{a}{2}   &  \cosh\frac{a}{2}
  \end{array}
  \right)\,
    \left(
  \begin{array}{cc}
    1+i\frac{b}{2} & \frac{b}{2} \\
    -\frac{b}{2} &  1-i\frac{b}{2}
  \end{array}
  \right)\, ,
\end{equation}
where $a,b$ are real variables and $-2\pi \leq \theta <2\pi$. Here the first
matrix is the SO(2) subgroup, the second is the Abelian factor A (dilatations,
or the subgroup SO(1,1)) and the third is the nilpotent factor N (corresponding
to conformal boosts). This decomposition allows us to define the {\it
compactified Minkowski space} as the boundary $S^1$ of the non-compact coset
${\cal D}^1=$K$\backslash$G. The group K acts transitively on $S^1$: The
stability subgroup of K at unity consists of two elements $I$ and $-I$, where
$I$ is the $2\times 2$ identity matrix. Thus, $S^1\sim$ K/$\{I,-I\}$. The
familiar Minkowski space coordinate $x\in \mathbb{R}^1$ can be introduced
analytically by using the Cayley transform: \bea e^{i\theta}=\frac{x-i}{x+i}.
\eea This construction works equally well in the case G = SU(2,2) and leads to
the definition of the compactified Minkowski space $\overline{\mathbb{M}}^{4}$
as a homogeneous space of the maximal compact subgroup K = SU(2)$\; \times\;
$SU(2)$\; \times\; $U(1).

Since we want to define a real space, it is preferable to work with the group
of real matrices SL(2,$\mathbb{R}$)$\sim$SU(1,1). In our specific
one-dimensional situation there is another possibility to introduce the
compactified Minkowski space. It is based on the Iwasawa decomposition G = NAK
of SL(2,$\mathbb{R}$):
\begin{equation}\label{29}
  g=\left(
  \begin{array}{cc}
    1 & 0 \\
    b & 1
  \end{array}
  \right)\, \left(
  \begin{array}{cc}
    e^{a/2} & 0 \\
    0 & e^{-a/2}
  \end{array}
  \right)\,  \left(
  \begin{array}{rr}
    \cos\theta & \sin\theta \\
    -\sin\theta & \cos\theta
  \end{array}
  \right)\,,
\end{equation}
where $a,b$ are real variables and $-\pi \leq \theta \leq \pi$. Here the first
matrix is the nilpotent factor $N$ (corresponding to conformal boosts), the
second is the Abelian factor $A$ (dilatations, or the subgroup SO(1,1) $
\subset$ SL(2,$\mathbb{R}$)) and the third is the compact factor K (the
subgroup SO(2) $ \subset$ SL(2,$\mathbb{R}$)). This decomposition allows us to
define the compactified Minkowski space as the coset NA$\backslash$G with the
additional requirements that the nilpotent factor N (the conformal boosts) acts
trivially on the fields. Then it is clear from eq. (\ref{29})  that the coset
representative is just the third factor, i.e., the group SO(2) or the unit
circle $S^1$. This time the Minkowski space coordinate $x\in \mathbb{R}^1$ can
be introduced analytically by changing from the angular variable $\theta$ on
the semicircle $-\pi/2 < \theta < \pi/2$ to the real variable $x = \tan\theta$,
$-\infty < x < \infty$. \footnote{The compact coset space NA$\backslash$G can
be restored by adding the infinite point to the real line $\mathbb{R}^1$.} We
stress again that identifying K with the compactified Minkowski space is
specific to SL(2,$\mathbb{R}$).

Alternatively, we can introduce Minkowski space through the Bruhat
decomposition NAM$\tilde{\mbox{N}}$ (see, e.g., \cite{DMPPT}):
\begin{equation}\label{30}
  g=\left(
  \begin{array}{cc}
    1 & b \\
    0 & 1
  \end{array}
  \right)\, \left(
  \begin{array}{cc}
    e^{a/2} & 0 \\
    0 & e^{-a/2}
  \end{array}
  \right)\,  (-1)^\epsilon\, \left(
  \begin{array}{cc}
    1 & 0 \\
    x & 1
  \end{array}
  \right)\,,
\end{equation}
where $\epsilon=0,1$,
for almost all matrices from SL(2,$\mathbb{R}$). The factor ${\rm
M}=\{\mathbb{I}, -\mathbb{I}\}$ in eq. (\ref{30}) is the center of the
group.
Then Minkowski space can be identified with the coset NAM$\backslash$G.
Thus,
the familiar coordinate $x$ parametrizes the right nilpotent factor in
(\ref{30}), which generates translations. The left nilpotent factor in
(\ref{30})  generates conformal boosts and is part of the so-called
``stability
subgroup" of the origin $x=0$. The rest of it is the Abelian factor A
(dilatations) and the factor M (if the space-time dimension is greater than
one, M becomes the Lorentz group SO$(d-1,1)$).

We recall the point of view on the conformal fields that we have adopted in
this article. We treat them as functions on the conformal group G
homogeneous
under the left action of some subgroup H. Thus, they effectively live on the
coset space H$\backslash$G. In the case at hand we have H = NAM. In CFT one
imposes the additional requirement that the boosts N are represented
trivially
on the physical fields, $T_{N}\Phi(0)=\Phi(0)$. This amounts to setting
$b=0$
in (\ref{30}). Then, ignoring the constant center element M, we see that our
conformal fields can be viewed as functions of the lower triangular matrices
\begin{equation}\label{32}
  g_0 = \left(
  \begin{array}{cc}
    e^{a/2} & 0 \\
    e^{-a/2} x & e^{-a/2}
  \end{array}
  \right)\,,
\end{equation}
homogeneous under the dilatations with parameter $a$. We remark that this
realization of Minkowski space is closely related to Dirac's projective cone
from Section \ref{Dirac}. Indeed,  the variable $a$ in (\ref{32})
corresponds
to the weight factor $\xi^1+\xi^2$, see (\ref{light3}) and (\ref{light4}).

The next step is to find a suitable  orthogonal basis for conformal fields with
manifest action of the dilatation group SO(1,1). This can be done by adapting
the discussion of Section \ref{u1} with some important modifications. The
details are given in Appendix A. Here we only present the result. Since SO(1,1)
is a non-compact group, the basis is labelled by a continuous complex variable
$\nu$. The basis vectors are
\begin{equation}\label{392'}
   t^\ell_+(0,\nu;a,x) =  C(\nu)\,
  e^{-a\ell}\,x^{\nu}_+ \,, \qquad  2\ell < \Re(\nu) < 0\,,
\end{equation}
restricted to positive values of $x$ only, and similar vectors $t^\ell_-$ for
$x<0$. Then a field of dilatation weight $\ell$ is given by the integral
expansion (Mellin transform)
\begin{equation}\label{41}
  \Phi^\ell(a,x) =  \frac{1}{2\pi i}\int^{c+i\infty}_{c-i\infty}  d\nu\,
  \left[b_+(\nu) t^\ell_+(0,\nu;a,x) +
  b_-(\nu) t^\ell_-(0,\nu;a,x)\right]  \equiv e^{-\ell a} \varphi(x)\,.
\end{equation}
The last term in this equation is the analog of eq. (\ref{light4}) from Dirac's
description of the conformal fields.

\section{Contraction of conformal fields}\label{contr}

In this section we discuss the main point of the paper: How to perform the
simultaneous contraction of the algebras of the conformal group SU(1,1) and the
R symmetry group SU(2) in terms of conformal fields. We show that, starting
from the very simple geometric idea of expanding the unit disc, one can derive,
step by step, all the ingredients of the abstract algebraic contraction
underlying the pp-wave limit.

\subsection{Expansion of the unit disc}

As explained in Section \ref{u1}, the free conformal fields can be defined on
the unit disc ${\cal D}^1$ parametrized by a complex coordinate $t$, $|t|<1$.
The geometric idea behind the pp-wave limit is very simple: We want to rescale
$t$, so that the radius of the disc becomes infinite and the disc covers the
whole complex plane $\mathbb{C}^1$. To this end we introduce the contraction
parameter $\Omega>0$:
\begin{equation}\label{42}
 t\ \rightarrow\ \Omega t \quad \Rightarrow \quad |t| < \frac{1}{\Omega}\,.
\end{equation}
Letting $\Omega\rightarrow0$, we achieve the desired result. So, we define the
 pp-wave limit as the ``expansion" of the space in which the conformal algebra is
realized.

Let us now see how this affects the conformal fields in the form (\ref{28}).
The weight $\ell$ in the non-compact case is negative, but in physics we are
used to positive weights, therefore we replace $-\ell=\Delta>0$. Then we
rewrite the basis vectors (\ref{27}) in the form
\begin{equation}\label{43}
  t^\Delta_{0n} \equiv e^{i\Delta\psi}\, f^\Delta_n(t,\bar t) =
  e^{i\Delta\psi} \frac{(-1)^n}{\sqrt{n!}} \left(\frac{\Gamma(2\Delta+n)}
 {{\Gamma(2\Delta)}}\right)^{1/2}\, (1-|t|^2)^\Delta {t^n}{}\,.
\end{equation}

We will take care of the phase factor $e^{i\Delta\psi}$  a bit later, for the
moment we concentrate on $f^\Delta_n(t,\bar t)$. Under the rescaling (\ref{42})
$t^n$ goes to zero like $\Omega^n$, so we ought to compensate this by an
appropriate behavior of the normalization factor in (\ref{43}). Setting
\begin{equation}\label{44}
  \Delta = \frac{\lambda}{2\Omega^2}\,,
\end{equation}
where $\lambda >0$ is a fixed constant, and using the asymptotic expressions
$$
\left(\frac{\Gamma({\lambda}/{\Omega^2}+n)}
 {{\Gamma({\lambda}/{\Omega^2})}}\right)^{1/2} \approx
 \frac{\lambda^{n/2}}{\Omega^n}\,, \qquad
 (1-\Omega^2 |t|^2)^{\lambda/2\Omega^2}  \approx e^{-\frac{1}{2}\lambda |t|^2}\,,
$$
valid for $\Omega\to 0$, we easily find
\begin{equation}\label{45}
  {\bf f}_n = \lim_{\Omega\to 0} f^\Delta_n = (-1)^n
  \sqrt{\frac{\lambda^n}{n!}}t^n e^{-\frac{1}{2}\lambda|t|^2}\,.
\end{equation}
Note that the constant $\lambda$ can be absorbed into the complex variable by the
rescaling $t\ \rightarrow \sqrt{\lambda} t$, so
\begin{equation}\label{45'}
  {\bf f}_n = \frac{(-1)^n}{\sqrt{n!}}t^n e^{-\frac{1}{2}|t|^2}\,.
\end{equation}
We conclude that what we have obtained in this well-defined contraction limit
is the familiar Fock-Bargmann basis for the Heisenberg  algebra.

The next question is what to do with the phase factor $e^{i\Delta\psi}$ in
(\ref{43}), which diverges when $\Delta \sim 1/\Omega^2$.
At this point we should remember that our conformal fields are defined as UIRs
of both the conformal and R symmetry groups. In other words, they can be
regarded as functions on SU(1,1)$\; \times\; $SU(2), homogeneous under the left
action of the subgroup U(1)$\; \times\; $U(1) or, equivalently, as functions on
${\cal D}^1\times S^2$. Combining eqs. (\ref{25}) and (\ref{28}) and denoting
the SU(2) weight $\ell=J>0$ (``isospin"), we can write down
\begin{equation}\label{46}
  \Phi^{\Delta,J}(t,\bar t,\psi; \tau, \bar\tau, \phi) =
  \sum_{n=0}^{\infty}\sum_{m=0}^{2J}a_{mn} t^\Delta_{0n}(t,\bar t,\psi) t^J_{0m}(\tau, \bar\tau,
  \phi)= e^{i(\Delta\psi - J\phi)}\varphi(t,\bar t,\tau, \bar\tau)\,.
\end{equation}
Now we see two distinct possibilities to render the phase factor convergent.
The first consists in rescaling each of two U(1) coordinates as follows:
\begin{equation}\label{1stopt}
  \psi = \Omega^2\tilde\psi\,, \qquad
 \phi = \Omega^2\tilde\phi\,,
\qquad
  \lim_{\Omega\to 0} e^{i\Delta\psi} = e^{i\lambda\tilde\psi}\, ,
 \qquad
 \lim_{\Omega\to 0} e^{iJ\phi} = e^{i\lambda'\tilde\phi}
\,.
\end{equation}
In principle, we could choose to contract the conformal algebra su(1,1) alone,
i.e., not to touch the R symmetry sector at all. However, to have a
well-defined contraction limit for the full superconformal algebra (including
odd generators) requires that the contraction procedure must be the same for
both algebras (see \cite{HKS}).

The new variables $\tilde\psi, \tilde\phi$ run in the interval
$-2\pi/\Omega^2\leq \tilde{\psi}, \tilde{\phi}< 2\pi/\Omega^2$. Therefore, in
the contraction limit they cease to be angular coordinates and become
coordinates on $\mathbb{R}$. At the same time, the exponential functions
(\ref{1stopt}) obtained in the contraction limit are naturally
$\delta$-function normalized. This switching from Kronecker to
$\delta$-function normalization is in fact the well-known procedure of passing
from a Fourier series to a Fourier integral expansion and it allows
$\lambda,\lambda'$ to take arbitrary (non-integer) values. In fact, taking the
limit in this way we convert su(1,1)$\;\oplus\;$su(2) into two copies of the
harmonic oscillator algebra h(2), for which $\lambda$ and $\lambda'$ are two
different {\it central charges} (See Section \ref{Heis}). Clearly, this algebra
does not contain the Hamiltonian, i.e., the external automorphism U(1)
generator. Moreover, if we take the same limit at the level of the
superconformal algebra whose bosonic subalgebra is su(1,1)$\;\oplus\;$su(2), we
would find out that the corresponding contracted superalgebra is not that of
the pp-wave limit \cite{BFHP2}. Therefore, we consider the contraction
possibility (\ref{1stopt}) as {\it trivial}.

The second option to make the phase factor in (\ref{46}) converge is to
compensate the growth of $\Delta$ by that of $J$. To achieve this we set
\begin{equation}\label{47}
  \Delta + J = \frac{\lambda}{\Omega^2}\,, \quad \Delta-J = \mu\,, \qquad
  \psi+\phi = 2\hat\psi \,, \quad \psi-\phi = 2\Omega^2 \hat\phi
\end{equation}
and obtain
\begin{equation}\label{48}
  \lim_{\Omega\to 0} e^{i(\Delta\psi - J\phi)} = e^{i(\mu\hat\psi + \lambda
  \hat\phi)}\,.
\end{equation}

This is not yet the end of the story: Forcing $J$ to grow, we may have trouble
with the normalization factor in (\ref{24}), unless we rescale the complex
coordinate $\tau$ on the sphere in exactly the same way as that on the disc
(see eq. (\ref{42})), $\tau\ \rightarrow\ \Omega\tau$. Then the phase
independent part of the basis vectors (\ref{24}) has the well-defined limit
\begin{equation}\label{45''}
  {\bf g}_n = \frac{1}{\sqrt{n!}}\tau^n e^{-\frac{1}{2}|\tau|^2}\,
\end{equation}
(as before, the constant $\lambda$ has been absorbed into the coordinate). We
recognize the basis for another copy of the Heisenberg algebra.

Combining all this, we finally obtain
\begin{equation}\label{49}
  \lim_{\Omega\to 0}\Phi^{\Delta,J}(t,\bar t,\psi; \tau, \bar\tau, \phi) =
e^{i(\mu\hat\psi + \lambda
  \hat\phi)}\, \sum_{m,n=0}^\infty c_{mn} t^m \tau^n e^{-\frac{1}{2}|t|^2-\frac{1}{2}|\tau|^2}\,.
\end{equation}
Note that according to Eq. (\ref{47}) the new variables $\hat{\psi}$ and
$\hat{\phi}$ run over the intervals $-2\pi \leq \hat{\psi} < 2\pi$ and
$-2\pi/\Omega^2\leq \hat{\phi}< 2\pi/\Omega^2$, respectively. Therefore, in the
contraction limit $\hat{\phi}$ becomes a central charge coordinate, while
$\hat{\psi}$ remains an angular U(1) coordinate. Thus, the effect of the
limiting procedure (\ref{47}) is to preserve the diagonal u(1) subalgebra in
su(1,1)$\; \oplus\; $su(2). Since this is what one expects from the contracted
superconformal algebra in the pp-wave context (see \cite{BFHP2,HKS}), we call
this limit {\it non-trivial}.

\subsection{The Heisenberg algebra h(2) as a contraction of
su(1,1)$\; \oplus\; $su(2)}\label{Heis} Let us now examine what happens to the
algebra su(1,1)$\; \oplus\; $su(2) in the trivial (\ref{1stopt}) or non-trivial
(\ref{47}) limits. Since in the case (\ref{1stopt}) the conformal and the R
symmetry algebras are contracted independently, it is enough to consider the
contraction of su(1,1). We now rescale $t$ and replace $\psi$ by $\tilde\psi$
not just in the fields, but directly in the generators (\ref{2500}) of the
algebra su(1,1). It is clear that all the generators diverge, unless we
introduce appropriate rescalings:
\begin{eqnarray}
  && {\cal Q} = \lim_{\Omega\to 0}\Omega^2M_0 =
  -i{\partial\over \partial\tilde\psi} \nonumber\\
  && {\cal M}_+= \lim_{\Omega\to 0}\Omega \sqrt{2} M_+ =
  i{\partial\over \partial t} - \frac{\bar t}{2}
  {\partial\over \partial\tilde\psi} \label{1stgen}\\
  && {\cal M}_-= \lim_{\Omega\to 0}\Omega \sqrt{2}M_- =
  i{\partial\over \partial \bar t} + \frac{t}{2}
  {\partial\over \partial\tilde\psi}  \nonumber
\end{eqnarray}
It is easy to check that these generators satisfy the following harmonic
oscillator algebra
\begin{eqnarray}
 &&[{\cal M}_{-},{\cal M}_{+}]={\cal Q}\, ,~~~~
 [{\cal Q}, {\cal M}_\pm]=0\, , \label{harmosc}
\end{eqnarray}
where ${\cal Q}$ is the central element and ${\cal M}_\pm$ are the creation and
annihilation operators. Note the absence of the Hamiltonian, i.e., of a U(1)
generator rotating  ${\cal M}_\pm$.

The second limiting procedure  (\ref{47}) leads to the following generators:
\begin{align}
  &{\cal H} = \lim_{\Omega\to 0}(M_0+ L_0) =-
    i{\partial\over \partial\hat\psi} \,
 && {\cal Q} = \lim_{\Omega\to 0}\Omega^2(M_0 - L_0) =
  -i{\partial\over \partial\hat\phi} \nonumber\\
&  {\cal L}_+= \lim_{\Omega\to 0}\Omega \sqrt{2}L_+ = e^{i\hat\psi}
  \left[i{\partial\over \partial \tau} - \frac{\bar\tau}{2} {\partial\over \partial\hat\phi}\right] \nonumber
   &&{\cal L}_-= \lim_{\Omega\to 0}\Omega \sqrt{2}L_- = e^{-i\hat\psi}
  \left[i{\partial\over \partial \bar\tau} + \frac{\tau}{2} {\partial\over \partial\hat\phi}\right] \nonumber
\\
 &  {\cal M}_+= \lim_{\Omega\to 0}\Omega \sqrt{2} M_+ = e^{i\hat\psi}
  \left[i{\partial\over \partial t} - \frac{\bar t}{2} {\partial\over \partial\hat\phi}\right]
&&{\cal M}_-= \lim_{\Omega\to 0}\Omega \sqrt{2}M_- = e^{-i\hat\psi}
  \left[i{\partial\over \partial \bar t} + \frac{t}{2} {\partial\over \partial\hat\phi}\right] \nonumber
\end{align}
\begin{equation}\label{50}\end{equation}
and algebra:
\begin{eqnarray}
 &&[{\cal M}_{-},{\cal M}_{+}]={\cal Q}\, ,~~~~ [{\cal L}_{-},{\cal
L}_{+}]={\cal Q}\, ,\nonumber\\
 && [{\cal H}, {\cal M}_\pm]=\pm  {\cal M}_\pm\,
,~~~ [{\cal H}, {\cal L}_\pm]=\pm  {\cal L}_\pm\, . \label{HW}
\end{eqnarray}
Clearly, ${\cal M}_\pm$, ${\cal L}_\pm,$ and ${\cal Q}$ generate the algebra
h(2), i.e. two copies of the Heisenberg algebra. The two algebras share the
same central element ${\cal Q}$ and external U(1) automorphism ${\cal H}$. Then
it becomes clear that the charge $\lambda$ of the conformal field (\ref{49})
associated with the coordinate $\hat\phi$ corresponds to the central charge
${\cal Q}$ in (\ref{HW}), while the charge $\mu$ associated with the coordinate
$\hat\psi$ corresponds to the U(1) automorphism generator ${\cal H}$. In
physics ${\cal H}$ is interpreted as the Hamiltonian $1/2(p^2+q^2)$. The
difference $\mu=\Delta-J$ can be viewed as the anomalous dimension of a CFT
operator with canonical dimension $\Delta_0=J$. In the dual string theory the
charge $\lambda$ is identified with the light-cone momentum $p_+$ of a free
string moving in the pp-wave background \cite{BMN}.

Let us summarize what we have done. We started from the simple geometric
postulate of expanding the unit disc, on which the one-dimensional conformal
fields live. To make the basis vectors for such fields survive, we had to let
their conformal dimension $\Delta$ grow like the disc radius squared. Then, to
compensate the divergence in the U(1) factor which carries the UIR weight we
had either to transform this U(1) charge into a central charge, or to seek help
from the R symmetry sector by combining two U(1) charges into one central
charge. In the latter case it was important that in both the conformal and R
symmetry sectors we could find similar U(1) factors. This explains why we had
to use a realization of the conformal fields with manifest action of SO(2)
rather than the more familiar dilatation SO(1,1). As a result, we found a
generalized In\"on\"u-Wigner contraction of the algebra su(1,1)$\; \oplus\;
$su(2). This contraction has been performed, at the abstract algebraic level,
in the four-dimensional case su(2,2)$\; \oplus\; $su(4) $\sim$ so(4,2)$\;
\oplus\; $so(6) in Ref. \cite{HKS} (we quote this result in Appendix B). There
it has also been shown that this type of contracted (super)algebras underlie
the ``pp-wave limit" proposal of Ref. \cite{BFHP1}.

Finally, we remark that in the four-dimensional case the corresponding
contraction procedure would lead to the Fock-Bargmann realization of the
Heisenberg algebra h(8) on functions over the complex space $\mathbb{C}^8$.
This representation is unitary equivalent, via the integral transform from the
Fock-Bargmann picture to the familiar (coordinate) Schr\"odinger representation
of h(8) acting on functions over the real space $\mathbb{R}^8$, {\it the latter
being the target space} of the dual light-cone string theory in the pp-wave
background \cite{M,MT}. It would be interesting to understand better the
relationship between the CFT and the corresponding dual supergravity (string)
theory from this point of view.

\subsection{Contraction of Minkowski fields}

One may wonder if it is possible to carry out the contraction procedure
directly for the conformal fields defined in the compactified Minkowski space
$\overline{\mathbb{M}}$. In our simplified situation $\overline{\mathbb{M}}^1 =
S^1$. It is known that the discrete series ${\cal D}_{\ell}^{\pm}$ of SU(1,1) can
be realized in the space of functions on the unit circle as well (see, e.g.,
Ref. \cite{Vilenkin}). In particular, the functions of ${\cal D}_{\ell}^{+}$
are expanded in a Fourier series, \bea \label{fex}
f(\theta)=\sum_{n=0}^{\infty}a_n e^{in\theta} \eea and may be viewed as
limiting (boundary) values of functions holomorphic inside the unit disc (cf.
(\ref{28})). The action of the infinitesimal generators (\ref{3}) is given by
\begin{eqnarray} \nonumber
&&M_+=-ie^{-i\theta} d/d\theta\, \\
\label{gencir}
&&M_-=ie^{-i\theta} d/d\theta-2\ell e^{i\theta}  \, \\
\nonumber &&M_0=\ell+i d/d\theta \, .
\end{eqnarray}
Note that in (\ref{fex}) the conformal field is expanded over the basis
${e^{in\theta}}$ in which the compact SO(2) generator $M_0$ is diagonal.

In Section \ref{Heis} we explained how to perform the pp-wave contraction at
the level of the Lie algebra. Now we have a different explicit realization of
the same algebra. Requiring the generators (\ref{gencir}) to stay non-singular
in the contraction limit, we have to try to find the corresponding contraction
prescription for the single coordinate of the carrier space $\theta$.
Considering $\theta$ as a function of $\Omega$ and demanding, for instance,
$\Omega^2M_0$ to be finite when $\Omega\to 0$, we deduce that either $\ell\sim
1/\Omega^2$ and $\theta \sim (a_1+a_2\Omega+a_3\Omega^2) $ or $\ell\sim
1/(a_1+a_2\Omega+a_3\Omega^2)$ and $\theta \sim \Omega^2$, where one of the
coefficients $a_i$ can be non-zero. In neither of these two cases a
non-singular limit for the generators $\Omega M_{\pm}$ can be obtained. In the case
of the expanding unit disc we had an additional radial coordinate whose
expansion rate was adjusted to compensate the contraction of the Lie algebra
generators. In the present situation the expansion of the unique angular
coordinate is not sufficient to compensate the different contraction rates of
the generators $M_{\pm}$ and $M_0$. Thus, we are lead to the conclusion that a
smooth pp-wave contraction limit for conformal fields defined in the
compactified Minkowski space is not possible.

Finally, let us recall the alternative realization of the conformal fields in
Minkowski space viewed as a coset of the group SL(2,$\mathbb{R}$) (Section
\ref{Mink}). The distinctive feature of this realization is that the fields
transform homogeneously under the non-compact dilatation group SO(1,1) instead
of the compact SO(2). This is reflected in the exponential factor $e^{-\ell a}
= e^{\Delta a}$ in eq. (\ref{41}). Trying to perform the contraction on such
fields, we have no choice but to rescale the SO(1,1) carrier coordinate $a$ and
thus to transform the weight $\Delta$ into a central charge. Indeed, the
compact R symmetry group {\it does not contain an SO(1,1) subgroup}, so it
cannot provide a compensating factor for the diverging non-compact factor
$e^{\Delta a}$. We are lead to the conclusion that when trying to perform the
contraction not just at the abstract algebraic level, but on the UIRs of the
algebra (i.e., on the fields) we have to carefully choose the space where these
UIRs are defined.

\section{Correlation functions in the contraction limit}

The central point in CFT is the study of correlation functions. From the
group-theoretical point of view, these are multi-point covariants of the
conformal group. As we have seen, the pp-wave limit is essentially a transition
from field theory to quantum mechanics, where ``a second-quantized field''
describing a multiparticle system degenerates into the wave function of ``a
first-quantized particle''. Therefore, the very notion of a correlation
function becomes rather obscure and the relevant question we may ask is: Can
the multi-point covariants of the conformal and R symmetry groups survive in
this quantum mechanical limit, where the conformal weight and the R charge
become infinite?

Here we propose a possible answer to this question.\footnote{The discussion in
this section is largely inspired by experience with harmonic superspace
\cite{Kniga,hh,Pickering}, where one systematically uses coordinates for the R
symmetry group on the same footing as the space-time coordinates.}

\subsection{Two-point functions}

Let us begin with the simplest case of the two-point functions. 
Consider first the conformal group SU(1,1) alone. A suitable
two-point covariant can be constructed very easily, remembering our
interpretation of the conformal fields as functions on G, covariant under the
left action of K but invariant under the right action of G. Let us take two
copies ${\bf g}_1,{\bf g}_2$ of the SU(1,1) matrices (\ref{26}) and denote
their entries by ${\bf g}_\pm{}^i$ (see Section \ref{leftrightsect}). Then it
is clear that the two-point functions
\begin{equation}\label{twpf}
 ({\bf 1_\pm 2_\pm}) \equiv ({\bf
g}_1)_\pm{}^i \epsilon_{ij} ({\bf g}_2)_\pm{}^j
\end{equation}
are invariant under the right action of the group (i.e., the action on the
indices $i,j$), and covariant under the left action (i.e., the action on the
indices $\pm$). In particular,
\begin{equation}\label{52}
  ({\bf 1_-2_-}) = e^{-i{(\psi_1+\psi_2)/2}} \frac{t_1-
t_2}{\sqrt{(1-|t_1|^2)(1-|t_2|^2)}}
\end{equation}
is a two-point covariant carrying weight $\Delta=-\ell=-1/2$ at each point,
while
\begin{equation}\label{522}
  ({\bf 1_-2_+})  = e^{i{(-\psi_1+\psi_2)/2}} \frac{t_1\bar t_2-1}{\sqrt{(1-|t_1|^2)(1-|t_2|^2)}}
\end{equation}
carries weight $\Delta_1=-1/2$ at point 1 and $\Delta_2=1/2$ at point
2.\footnote{The modulus of expression (\ref{522}) is the ``invariant distance" between
points 1 and 2 in the Riemannian sense.}

Two-point functions with equal positive weights $\Delta>0$ can be obtained as
follows:
\begin{equation}\label{53}
  \langle \Phi^\Delta| \Phi^\Delta\rangle_{SU(1,1)} =
  \frac{1}{({\bf 1_-2_-})^{2\Delta}} =
  e^{i\Delta{(\psi_1+\psi_2)}}
\frac{(1-|t_1|^2)^{\Delta}(1-|t_2|^2)^{\Delta}}{(t_1-t_2)^{2\Delta}}\,.
\end{equation}

In accordance with our interpretation of the fields $\Phi^\ell$ as lowest
weight vectors of the left algebra (\ref{2501}) (see (\ref{leftlws'})), we
obtain
\begin{eqnarray}
  && M_0\langle \Phi^\Delta| \Phi^\Delta\rangle =
  \langle \Phi^\Delta| \Phi^\Delta\rangle \overleftarrow{M}_0 =
  \Delta\ \langle \Phi^\Delta| \Phi^\Delta\rangle\,, \label{accord1}\\
  && M_-\langle \Phi^\Delta| \Phi^\Delta\rangle =
  \langle \Phi^\Delta| \Phi^\Delta\rangle \overleftarrow{M}_- = 0\,.\label{accord2}
\end{eqnarray}
It should be stressed that when checking eq. (\ref{accord2}) we have neglected
the singularity at the coincident point $t_1=t_2$, which is a typical property
of the two-point functions of physical fields (in reality the right-hand side
of eq. (\ref{accord2}) contains a contact term).

Note that there exists another right-invariant expression with the same weight
$\Delta$, $({\bf 1_+2_+})^{2\Delta}$. It is not a suitable two-point function,
because it is annihilated by $M_+$ instead of $M_-$, so it corresponds to a
highest weight vector of {\it positive} weight. As a consequence, the explicit
expression of $({\bf 1_+2_+})^{2\Delta}$ contains the antiholomorphic
polynomial $(\bar t_1 - \bar t_2)^{2\Delta}$. This means that we are dealing
with a finite-dimensional non-unitary representation of SU(1,1).

The two-point covariant (\ref{522}) is of a different nature. First of all,
since the variables $t_{1,2}$ are restricted to the interior of the unit disc,
$|t_{1,2}|<1$, this expression never vanishes. Then the two-point function
\begin{equation}\label{53'}
  \langle \Phi^\Delta| \bar\Phi^{-\Delta}\rangle_{SU(1,1)} =
  \frac{(-1)^{2\Delta}}{({\bf 1_-2_+})^{2\Delta}} =
  e^{i\Delta{(\psi_1-\psi_2)}}\
  \frac{(1-|t_1|^2)^\Delta (1-|t_2|^2)^\Delta}{(1-t_1\bar t_2)^{2\Delta}}
\end{equation}
is {\it non-singular at the coincident point}, unlike (\ref{53}). Secondly, eq.
(\ref{accord2}) is replaced by
\begin{equation}\label{accord3}
  M_-\langle \Phi^\Delta| \bar\Phi^{-\Delta}\rangle =
  \langle \Phi^\Delta| \bar\Phi^{-\Delta}\rangle \overleftarrow{M}_+ = 0\,.
\end{equation}
This means that the two-point function (\ref{53'}) corresponds to a
lowest-weight irrep (holomorphic) of {\it positive} weight $\Delta$ at point 1
and to a highest-weight irrep (antiholomorphic) of {\it negative} weight
$-\Delta$ at point 2. In fact, this is the so-called Bergman kernel
\cite{Berg, Ruhl} used to project the Hilbert space of functions square
integrable inside the unite disc, onto holomorphic functions on the disc.
Applied to a holomorphic field, this projector is the identity; in other words,
(\ref{53'}) plays the r\^ole of a delta function for holomorphic fields:
\begin{equation}\label{delhol}
  \Phi^\Delta(1) = \int_2 \langle \Phi^\Delta(1)| \bar\Phi^{-\Delta}(2)\rangle_{SU(1,1)} \
  \Phi^\Delta(2)\,.
\end{equation}

We remark that our two-point functions can also be regarded as functions on two
copies of the homogenous space ${\cal D}^1 \sim$ U(1)$\backslash$SU(1,1). To
this end we should drop the phase factor in (\ref{53}) or (\ref{53'}) (recall
(\ref{28})) and examine the transformation properties of the resulting function
of $t_1,t_2$ under the right group (generators (\ref{discgen})). Thus we make
contact with the more familiar interpretation of the two-point functions.

\subsection{Contraction of two-point functions}

Let us start with the {\it non-singular} two-point function (\ref{53'}). Using
the prescriptions (\ref{42}), (\ref{44}) and (\ref{1stopt}), we easily obtain
\begin{equation}\label{nonslim}
  \lim_{\Omega\to0}\langle \Phi^\Delta| \bar\Phi^{-\Delta}\rangle_{SU(1,1)}
 =e^{i\lambda{(\tilde\psi_1-\tilde\psi_2)}}e^{\lambda \left(t_1\bar{t}_2-\frac{1}{2}|t_1|^2
-\frac{1}{2}|t_2|^2 \right)   } \, .
\end{equation}
It can be shown that after a transformation from the Fock-Bargmann
to the Schr\"odinger (coordinate) realization of the Heisenberg algebra the
expression (\ref{nonslim}) becomes just a delta function, in accord with the
r\^ole of the function (\ref{53'}) before the contraction (recall
(\ref{delhol})).

It is important to realize that the limit (\ref{nonslim}) has been obtained
without the participation of the R symmetry sector. Consequently, the
contracted two-point function carries no information about the ``anomalous
dimension" $\mu=\Delta-J$, the only quantum number of interest in the pp-wave
limit. We note that this result corresponds to the two-point functions for the
Heisenberg algebra proposed in \cite{KP}.

Let us now turn to the {\it singular} two-point function (\ref{53}). Trying to
contract it, we run into two problems. Apart from the oscillating phase factor,
we have to cope with the infinite power in $(t_1-t_2)^{-2\Delta}$. As in
Section \ref{Heis}, we can seek help from the R symmetry sector. Indeed, our
fields are representations not only of the conformal group SU(1,1), but also of
the R symmetry group SU(2). Thus, we have to construct two-point functions
involving the SU(2) coordinates (\ref{23}) as well. This can be done as above,
by taking two SU(2) matrices $g_1,g_2$ and forming the right-invariant
expression
\begin{equation}\label{52'}
  ({1_-2_-}) \equiv (g_1)_-{}^i
\epsilon_{ij} (g_2)_-{}^j = e^{-i{(\phi_1+\phi_2)/2}} \frac{\tau_1-
\tau_2}{\sqrt{(1+|\tau_1|^2)(1+|\tau_2|^2)}}
\end{equation}
which carries weight $J=\ell=1/2$ at both points. For arbitrary weight $J$ we
can write down the two-point SU(2) covariant
\begin{equation}\label{54}
  \langle J| J\rangle_{SU(2)} = (1_-2_-)^{2J} =
  e^{-iJ{(\phi_1+\phi_2)}}
\frac{(\tau_1-\tau_2)^{2J}}{(1+|\tau_1|^2)^{J}(1+|\tau_2|^2)^{J}}\,.
\end{equation}
It satisfies equations similar to (\ref{accord1}), (\ref{accord2}). The main
difference from the SU(1,1) case is that the two-point function (\ref{53}) is
singular at the coincident point, while (\ref{54}) vanishes there. This is
explained by the fact that SU(2) has finite-dimensional UIRs which are given by
polynomials in the group coordinates $\tau$, see eq. (\ref{25}).

The next step is to multiply the two expressions (\ref{53})  and (\ref{54}) to
obtain the two-point function of fields with conformal weight $\Delta$ and
isospin $J$:
\begin{eqnarray}
  \langle \Phi^{\Delta, J}| \Phi^{\Delta, J}\rangle &=&
  \frac{\Gamma(J+1)}{\Gamma(\Delta+1)}
 \ \frac{(1_-2_-)^{2J}}{({\bf 1_-2_-})^{2\Delta}} \label{55} \\
  &=& \frac{\Gamma(J+1)}{\Gamma(\Delta+1)}
   e^{i\Delta{(\psi_1+\psi_2)}-iJ{(\phi_1+\phi_2)}}\,
   \frac{(1-|t_1|^2)^{\Delta}(1-|t_2|^2)^{\Delta}}
   {(1+|\tau_1|^2)^{J}(1+|\tau_2|^2)^{J}}\,
   \frac{(\tau_1-\tau_2)^{2J}}{(t_1-t_2)^{2\Delta}} \,.   \nonumber
\end{eqnarray}
The choice of normalization is justified by the contraction limit below.

We already know that the phase factor is well behaved in the limit (\ref{47}).
However, the last factor in (\ref{55}) is still divergent, unless we impose a
condition relating the space-time and R symmetry variables, for instance,
\begin{equation}\label{56'}
  (t_1-t_2)^2 = (\tau_1-\tau_2)^2
\end{equation}
or even stronger,
\begin{equation}\label{56}
 t_1=\tau_1 \quad {\rm and} \quad t_2=\tau_2\,.
\end{equation}
This allows us to find the contraction limit
\begin{equation}\label{57}
  \lim_{\Omega\to 0} \langle \Phi^{\Delta, J}| \Phi^{\Delta, J}\rangle =
\left(\frac{2}{\lambda}\right)^{\mu} e^{i\mu(\hat{\psi}_1+\hat{\psi}_2)}\
e^{i\lambda[(\hat{\phi}_1 + i|t_1|^2) + (\hat{\phi}_2 +i|t_2|^2)]} \
\frac{1}{(t_1-t_2)^{2\mu}}\, .
\end{equation}

We remark that before the contraction the coordinates $t$ and $\tau$
parametrize different spaces, the unit disc ${\cal D}^1$ and the two-sphere
$S^2$, respectively. However, the former is a subspace of the latter, as ${\cal
D}^1$ can be identified with, e.g., the southern hemisphere of $S^2$. In the
contraction process the unit disc as well as the southern hemisphere expand and
cover the whole complex plane. This allows us to perform the identification
(\ref{56}).

The meaning of eq. (\ref{56}) is that the contracted two-point functions of the
singular type can make sense only along certain hypersurfaces in the extended
space $\mathbb{C}^2$ where the Heisenberg algebra h(2) (\ref{HW}) acts. In such
a restricted subspace only half of the algebra h(2) is realized, as can be
seen, e.g., by the fact that the function (\ref{57}) is annihilated only by the
sum of the left ``annihilators" ${\cal L}_- + {\cal M}_-$. Alternatively,
introducing new coordinates $q_{\pm}=1/2(t \pm \tau)$ as well as new generators
${\cal A}_{\pm}=1/\sqrt{2}({\cal L}_{\pm}+{\cal M}_{\pm})$, one can check that ${\cal A}_{\pm}$
together with ${\cal Q}$ and ${\cal H}$ from eq. (\ref{50}) obey the
commutation relations of the algebra h(1) on the space of functions independent
of the coordinate $q_-$.

It is important to realize that in the contraction limit of the singular
two-point function the information about the anomalous dimension $\mu$ is not
lost, it is explicitly present in the coordinate factor $1/(t_1-t_2)^{2\mu}$.
This, in our opinion, makes meaningful the computation of the anomalous
dimension in the pp-wave limit by QFT methods.

\subsection{Three- and four-point functions}

Once we have defined the two-point functions, we can easily construct
multi-point functions as well. The correlation function of three fields
$\Phi^{\Delta_i,J_i}(t_i,\tau_i)$, $i=1,2,3$ is obtained by simply multiplying
the two-point ones:
\begin{equation}\label{3pt}
  \langle 1|2|3\rangle =
  C_{123}\prod_{i=1}^3\frac{\Gamma(\alpha_i/2 +1)}
  {\Gamma(\beta_i/2 +1)}\
  \frac{(12)^{\alpha_1}(23)^{\alpha_2}(31)^{\alpha_3}}
  {({\bf 12})^{\beta_1}({\bf 23})^{\beta_2}({\bf 31})^{\beta_3}} \,,
\end{equation}
where
\begin{equation}\label{albe}
  \alpha_i = J_i-J_j-J_k\,, \qquad
  \beta_i = \Delta_i-\Delta_j-\Delta_k\,, \qquad   i\neq j \neq k
\end{equation}
and we have used the short-hand notation $(12)\equiv (1_-2_-)$, etc. The
normalization factors $\Gamma(\alpha_i/2 +1)/\Gamma(\beta_i/2 +1)$ are needed
for convergence in the contraction limit. The finite normalization $C_{123}$ is
to be determined by the dynamics. The contraction can be performed after
identifying the space-time and R symmetry coordinates at each point (recall
(\ref{56})). The resulting expression contains three central charges
$\lambda_i$ and anomalous dimensions $\mu_i$.

In the four-point case, for simplicity we  consider four identical fields
$\Phi^{\Delta,J}$. The most general four-point function then is
\begin{equation}\label{4pt'}
  \langle 1|2|3|4 \rangle_{\Delta,J} =
  \frac{\Gamma^2(J +1)}
  {\Gamma^2(\Delta +1)}\
  \frac{[(12)(34)]^{2J}}
  {[({\bf 12})({\bf 34})]^{2\Delta}}\ \sum_{n=0}^{2J} U^n f_n(u) \,.
\end{equation}
Here the covariant prefactor is a product of two-point functions with the
corresponding weights. In addition, we have introduced a dependence on the
conformal and R symmetry invariants (``cross-ratios") $u,U$. In our
one-dimensional case there is only one such four-point invariant of each kind,
which is easily obtained \cite{Pickering} by combining two-point covariants of
the type (\ref{52}), e.g.,
\begin{equation}\label{crrat}
  u = \frac{({\bf 13})({\bf 24})}{({\bf 12})({\bf 34})} \quad \mbox{and} \quad U = \frac{({ 13})({ 24})}{({ 12})({
  34})}\,.
\end{equation}
Note that the alternative choice $({\bf 14})({\bf 23})$ for the numerator is
equivalent due to the cycling identity
\begin{equation}\label{cycl}
  ({\bf 12})({\bf 34}) + ({\bf 13})({\bf 42}) + ({\bf 14})({\bf 23}) = 0\,.
\end{equation}

The reason why the dependence on $U$ in eq. (\ref{4pt'}) must be polynomial of
order not greater than $2J$ is that the fields $\Phi^{\Delta,J}$ belong to a
finite-dimensional UIR of SU(2). Therefore the dependence on the R symmetry
coordinates $\tau_i$ at each point must be polynomial (recall (\ref{25})). In
other words, the singularity coming from the denominator in $U^n$ should not
exceed the power of $(12)(34)$ in the prefactor in eq. (\ref{4pt'}). The
dependence on the conformal cross-ratio $u$ is virtually arbitrary.

Finally, the contraction limit in (\ref{4pt'}) can be taken as before. The
prefactor is well-behaved owing to the $\Gamma$ factors and to the
identification of the two spaces. The cross-ratios themselves are non-singular
in the limit $\Omega\to0$. We stress that in this limit the polynomial
expansion in $U$ in (\ref{4pt'}) becomes an {\it infinite power series}. This
is clearly due to the fact that the tensor product of two R symmetry irreps
with infinite $J$ contains an infinite number of terms. At the same time,
because of the identification of the space-time and R symmetry coordinates, the
resulting four-point correlator will depend on an arbitrary function of a
single cross-ratio. This function should, in principle, be determined by the
dynamics of the theory.

\section*{Acknowledgements} We are grateful to V. Dobrev, E. Ivanov, S. {}Ferrara,
 S. {}Frolov, E. Kiritsis, S. Kuzenko, V. Petkova, R. Stora and S. Theisen for many useful discussions.
G. A. was supported by the DFG and by the
European Commission RTN programme HPRN-CT-2000-00131,
and in part by RFBI grant N99-01-00166.

\section{Appendix A}

Here we find the orthogonal basis for conformal fields with manifest action of
the dilatations, mentioned in Section \ref{Mink}. We adapt the discussion of
Section \ref{u1} with some important modifications. First of all, since we are
dealing with real matrices from SL(2,$\mathbb{R}$), we can replace the complex
spinor $z_1,z_2$ by a real one, $\xi_1,\xi_2 \in \mathbb{R}$. Next, we replace
the representation basis (\ref{13}) by
\begin{equation}\label{131}
  {\rm SL}(2,\mathbb{R}): \qquad |\ell,\mu\rangle = N(\mu)\, \xi_1^\mu
  \xi_2^{2\ell-\mu}\,,
\end{equation}
where $\mu\in \mathbb{C}$ (the reason will become clear in a moment). As
before, the group SL(2,$\mathbb{R}$) acts on the spinor $\xi_1,\xi_2$ by linear
transformations (cf. (\ref{16})). For the particular choice (\ref{32}) this
gives
\begin{equation}\label{171}
  T^\ell(g)|\ell,\mu\rangle  = N(\mu)\, e^{a(\mu-\ell)} \xi_1^{2\ell}
  (x+\xi)^{2\ell-\mu}\,,
\end{equation}
where $\xi = \xi_2/\xi_1$ and we have assumed that $\xi_1\neq 0$. Now, we have
to expand the right-hand side in the basis (\ref{131}). However, we cannot
proceed as we did after eq. (\ref{21}). The reason is that this time the
variable $x$ is not restricted to a finite domain, unlike the unit disc
variable $t$, $|t|<1$ from (\ref{26}). Thus, in (\ref{171}) we cannot achieve,
e,g., $|x/\xi|<1$ keeping $x$ and $\xi$ independent. This problem is a
manifestation of the fundamental difference between coset spaces with compact
and non-compact stability groups.

The correct expansion in the non-compact case is obtained through a Mellin
transform (a generalization of the Fourier transform). We recall that a
function $f(\xi)$, $\xi>0$ such that $\xi^{c_1-1}f(\xi)$ and
$\xi^{-c_2-1}f(\xi)$ are integrable for some $c_1,c_2>0$, admits the integral
transform
\begin{equation}\label{34}
  F(\lambda) = \int^\infty_0 f(\xi) \xi^{\lambda-1} d\xi
\end{equation}
with inverse
\begin{equation}\label{35}
  f(\xi) = \frac{1}{2\pi i}\int^{c+i\infty}_{c-i\infty}\,
   F(\lambda) \xi^{-\lambda} d\lambda
\end{equation}
for any $-c_2<c<c_1$. So, the matrix elements in the basis (\ref{131}) will be
determined by the integral expansion
\begin{equation}\label{351}
  T^\ell(g)|\ell,\mu\rangle  =  \frac{1}{2\pi i}\int^{c+i\infty}_{c-i\infty}\,
   t^\ell(\mu,\nu;g) |\ell,\nu\rangle d\nu\,.
\end{equation}

Let us now apply the Mellin transform (\ref{34}) to the factor
$(x+\xi)^{2\ell-\mu}$ in the right-hand side of eq. (\ref{171}) (for $\xi>0$):
\begin{equation}\label{37}
  F(\lambda) = \int^\infty_0 (x+\xi)^{2\ell-\mu} \xi^{\lambda-1} d\xi\,.
\end{equation}
Restricting the coordinate $x$ to positive values only, we obtain
\begin{equation}\label{38}
  F(\lambda) =  x^{2\ell-\mu+\lambda}_+\int^\infty_0 (\rho+1)^{2\ell-\mu}
  \rho^{\lambda-1} d\rho =
  \frac{\Gamma(\lambda)\Gamma(-2\ell+\mu-\lambda)}{\Gamma(\mu-2\ell)}\,
  x^{2\ell-\mu+\lambda}_+\,,
\end{equation}
where $\rho=\xi/x$ and the notation $x_+$ is a reminder that this expansion is
valid only for $x>0$.\footnote{The case $x<0$ can be treated similarly.} The
convergence conditions in eq. (\ref{38}) are
\begin{equation}\label{39}
  0 < \Re(\lambda) < \Re(\mu-2\ell)\,.
\end{equation}

Finally,  setting $\lambda=\nu-2\ell$ in eq.  (\ref{38}) and comparing it with
eqs. (\ref{131}), (\ref{171}) and (\ref{351}), we find the matrix
elements\footnote{We do need the explicit form of the normalization
coefficients $N(\mu)$.}
\begin{equation}\label{391}
  t^\ell_+(\mu,\nu;a,x) =  \frac{N(\mu)}{N(\nu)}
  \frac{\Gamma(\nu-2\ell)\Gamma(\mu-\nu)}{\Gamma(\mu-2\ell)}\, e^{a(\mu-\ell)}\,
  x^{\nu-\mu}_+\,, \qquad  2\ell < \Re(\nu) < \Re(\mu) \,.
\end{equation}

So, the relevant basis vectors for expanding conformal fields of weight $\ell$
in a {\it manifestly dilatation covariant way} are the functions
\begin{equation}\label{392}
   t^\ell_+(0,\nu;a,x) =  C(\nu)\,
  e^{-a\ell}\,x^{\nu}_+ \,, \qquad  2\ell < \Re(\nu) < 0
\end{equation}
(recall that $\ell<0$ for UIRs from the discrete series), and their analogs
$t^\ell_-(0,\nu;a,x)$ for $x<0$. They form an orthogonal set, e.g.
\begin{equation}\label{40}
  \int^\infty_0 \overline{x^{\nu'}_+}\, x^{\nu}_+  dx =2\pi i\,
  \delta(\nu +\overline{\nu'}+1)
\end{equation}
(this property is used to inverse the Mellin transform).


                             \section{Appendix B}

The bosonic part of the symmetry group of the ${\cal N}=4$ SYM is SU(2,2)$\;
\times\; $SU(4), which is locally isomorphic to SO(4,2)$\; \times\; $SO(6). In
the pp-wave limit this group (the Lie algebra) undergoes a generalized
In\"on\"u-Wigner contraction. It is useful to recall here the contraction
procedure for the corresponding Lie algebra. Our presentation is similar to
that of Ref. \cite{HKS}.

The Lie algebra so(4,2) with antisymmetric generators $M_{ab}$,
$M_{ab}=-M_{ba}$, has the following commutation relations \begin{eqnarray}
\label{Bcr}
[M_{ab},M_{cd}]=\eta_{bc}M_{ad}+\eta_{ad}M_{bc}-\eta_{ac}M_{bd}-\eta_{bd}M_{ac}\,
, \end{eqnarray} where $a,b=0,\ldots, 5$ and the metric is $\eta={\rm diag
}(-,+,...,+,-)$. It is useful to rewrite these relations by splitting the
generators $M_{ab}$ as $(M_{0i},M_{i5},M_{ij},M_{05})$ with $i,j=1,...,4$:
\begin{center}
\begin{eqnarray}
\begin{tabular}{lll}
$[M_{0i},M_{0j}]=M_{ij}$\,  & &     $[M_{i5},M_{j5}]=M_{ij}$ \,\\
$[M_{0i},M_{j5}]=\delta_{ij}M_{05}$\,  & &  $[M_{i5},M_{kl}]=\delta_{ik}M_{l5}-\delta_{il}M_{k5}$     \, \\
$[M_{0i},M_{kl}]=\delta_{ik}M_{0l}-\delta_{il}M_{0k}$\,  & &
\end{tabular}
\label{Bcomrel}
\end{eqnarray}
\end{center}
and
\begin{eqnarray}
\nonumber
&&[M_{05},M_{i5}]=M_{0i} \, \\
&&[M_{05},M_{0i}]=-M_{i5} \, \label{Bcrc}\\
\nonumber &&[M_{05},M_{kl}]=0 \,
\end{eqnarray}
The generators $M_{ij}$ obey the commutation relations of the Lie algebra
so(4). From (\ref{Bcomrel}) one concludes that the vectors $M_{0i}$ and
$M_{i5}$ transform in the vector representation of so(4). The generator
$M_{05}$ rotates $M_{i5}$ into $M_{0i}$ and $M_{0i}$ into $-M_{i5}$, so it is
the generator of SO(2). It is worthwhile to note that since the algebra so(4,2)
is defined over $\mathbb{R}$ one is not able to diagonalize $M_{05}$ in the
vector space spanned by $M_{0i}$ and $M_{i5}$.\footnote{Recall that the
generators of translations $P_{\mu}$, special conformal transformations
$K_{\mu}$ and dilatations $D$ are expressed via $M_{ab}$ as
$P_{\mu}=M_{\mu 5}-M_{\mu 4}$
$K_{\mu}=M_{\mu 4}+M_{\mu 5}$ and $D=M_{54}$ with $\mu=0,..,3$.
Thus, the SO(2) generator (usually referred to as the conformal Hamiltonian) is
$M_{05}=\frac{1}{2}(P_0+K_0)$.} Yet another convenient set of generators is
obtained by introducing $M_0=iM_{05}$, $M^{\pm}_i={i}/{\sqrt{2}}(M_{i5}\pm
iM_{0i})$. The corresponding commutation relations are \begin{eqnarray}
\nonumber
&&[M_0, M^{\pm}_i]=\pm M^{\pm}_i\, ,~~~~[M^+_i,M^-_j]=-\delta_{ij}M_0-M_{ij}\,  \\
\label{BC1} &&[M^{\pm}_i,M_{kl}]=\delta_{ik}M^{\pm}_l-\delta_{il}M^{\pm}_k \, .
\end{eqnarray}

The Lie algebra so(6) with generators $L_{ab}$, $a,b=1,..., 6$ has the same
relations (\ref{Bcr}), except that the metric is now $\eta_{ab}=\delta_{ab}$.
We rewrite the commutation relations by splitting the set of generators into
$L_{ij},L_{i5},L_{i6},L_{56}$ with $i,j=1,..,4$:
\begin{center}
\begin{eqnarray}
\begin{tabular}{lll}
$[L_{i5},L_{j5}]=-L_{ij}$\,   & &     $[L_{i6},L_{j6}]=-L_{ij}$ \, \\
$[L_{i5},L_{j6}]=-\delta_{ij}L_{56}$\,   & & $[L_{i6},L_{kl}]=\delta_{ik}L_{l6}-\delta_{il}L_{k6}$     \, \\
$[L_{i5},L_{kl}]=\delta_{ik}L_{l5}-\delta_{il}L_{k5}$\, , & &
\end{tabular}
\label{Bcomrelso}
\end{eqnarray}
\end{center}
and
\begin{eqnarray}
\nonumber
&&[L_{56},L_{i5}]=-L_{i6} \, ,\\
&&[L_{56},L_{i6}]=L_{i5} \, ,\label{Bcrcso}\\
\nonumber &&[L_{56},L_{kl}]=0 \, .
\end{eqnarray}
Here $L_{ij}$ generates the subalgebra so(4) and $L_{56}$ is the generator of
so(2). Like in the previous case, one can define $L_0=-iL_{56}$ and
$L^{\pm}_i={i}/{\sqrt{2}}(L_{i5}\pm iL_{i6})$ with commutation
relations\footnote{The generators have the Hermitian conjugation property
$L_0^{\dagger}=L_0$, $(L^{\pm})^{\dagger}=L^{\mp}$, and similarly for $M_0,M^{\pm}$.}
\begin{eqnarray}
\nonumber
&&[L_0, L^{\pm}_i]=\pm L^{\pm}_i\, ~~~~[L^+_i,L^-_j]=\delta_{ij}L_0+L_{ij}\,  \\
\label{BC2} &&[L^{\pm}_i,L_{kl}]=\delta_{ik}L^{\pm}_l-\delta_{il}L^{\pm}_k \, .
\end{eqnarray}

To perform the pp-wave contraction one introduces a parameter $\Omega$ and
rescales the generators as
\begin{eqnarray}
{\cal Q}=\Omega^2(M_0-L_0)\, ,~~~~ {\cal H}=M_0+L_0\, ,~~~~ {\cal M}^{\pm}_i=\Omega
\sqrt{2}M^{\pm}_i \, ,~~~~ {\cal L}^{\pm}_i=\Omega \sqrt{2}L^{\pm}_i \, ,
\end{eqnarray}
while the so(4) generators $L_{ij}={\cal L}_{ij}$ and $M_{ij}={\cal M}_{ij}$
are kept intact. The commutation relations for the new set of generators become
$\Omega$-dependent and in the limit $\Omega\to 0$ one finds the algebra
\begin{eqnarray}
\nonumber &&[{\cal M}_i^{-},{\cal M}^{+}_j]=\delta_{ij}{\cal Q}\, ,~~~~ [{\cal
L}_i^{-},{\cal L}^{+}_j]=\delta_{ij}{\cal Q}\, ,~~~~ [{\cal H}, {\cal
M}^{\pm}_i]=\pm  {\cal M}^{\pm}_i\, ,~~~
[{\cal H}, {\cal L}^{\pm}_i]=\pm  {\cal L}^{\pm}_i\, , \\
\label{BHW} &&[{\cal M}^{\pm}_i,{\cal M}_{kl}]=\delta_{ik}{\cal
M}^{\pm}_l-\delta_{il}{\cal M}^{\pm}_k\, ,~~~~ [{\cal L}^{\pm}_i,{\cal
L}_{kl}]=\delta_{ik}{\cal L}^{\pm}_l-\delta_{il}{\cal L}^{\pm}_k\,
\end{eqnarray}
with ${\cal Q}$ being a central element. Clearly, ${\cal M}_i^{\pm},{\cal L}_i^{\pm}$
generate the algebra h(8), i.e. eight copies of the Heisenberg algebra with the
common central element ${\cal Q}$, while ${\cal H}, {\cal M}_{ij}, {\cal
L}_{ij}$ generate its external automorphism algebra ${\rm so}(2)\; \oplus\;
$so(4)$\; \oplus\; $so(4).

\end{document}